\documentclass[iop,revtex4]{emulateapj}

\usepackage{amsmath}
\usepackage{amssymb}
\usepackage{color}
\usepackage{epic}
\usepackage{epstopdf}
\usepackage{graphicx}
\usepackage{hyperref}
\usepackage{multirow}
\usepackage{slashbox}
\usepackage{natbib}
\usepackage{overpic}
\usepackage{soul}
\usepackage{textcomp}
\usepackage{lscape}
\usepackage[figuresright]{rotating}

\shorttitle{NGC~253 at low radio frequencies}
\shortauthors{\textsc{Kapi\'nska et al.}}

\begin{document}

\title{Spectral energy distribution and radio halo of NGC~253 at low radio frequencies}

\author{A.~D. Kapi\'{n}ska$^{1,2}$, L.~Staveley-Smith$^{1,2}$, R. Crocker$^{3}$, G. R. Meurer$^1$, S. Bhandari$^{4,2}$, N.~Hurley-Walker$^{5}$, A.R.~Offringa$^{6}$, D.J. Hanish, N. Seymour$^5$, R.~D. Ekers$^7$, 
M.~E.~Bell$^{7}$, J.R.~Callingham$^{7,8,2}$, K.~S.~Dwarakanath$^{9}$, B.-Q.~For$^{1}$, B.~M.~Gaensler$^{10,8,2}$, P.~J.~Hancock$^{5,2}$, L.~Hindson$^{11,12}$, M.~Johnston-Hollitt$^{12}$, E.~Lenc$^{8,2}$, B.~McKinley$^{13}$, J.~Morgan$^{5}$, P.~Procopio$^{13,2}$, R.~B.~Wayth$^{5,2}$, C.~Wu$^1$, Q.~Zheng$^{12}$, 
N. Barry$^{14}$,  A.~P. Beardsley$^{14}$, J.~D. Bowman$^{15}$,  F. Briggs$^3$, P. Carroll$^{15}$, J.~S. Dillon$^{16}$, A. Ewall-Wice$^{16}$, L. Feng$^{16}$, L.~J. Greenhill$^{17}$, B.~J. Hazelton$^{14}$, J.~N.  Hewitt$^{16}$, D.~J. Jacobs$^{15}$, H.-S. Kim$^{13,2}$, P. Kittiwisit$^{15}$, J. Line$^{13,2}$, A. Loeb$^{17}$, D.~A. Mitchell$^{3,2}$, M.~F. Morales$^{14}$, A.~R. Neben$^{16}$, S. Paul$^{9}$, B. Pindor$^{13,2}$, J.~C. Pober$^{18}$, J. Riding$^{13,2}$, S.~K. Sethi$^{9}$, N.~Udaya~Shankar$^{9}$, R.~Subrahmanyan$^{9,2}$,  I.~S. Sullivan$^{14}$, M. Tegmark$^{16}$, N. Thyagarajan$^{15}$, S.~J.~Tingay$^{5,19,2}$, C.~M. Trott$^{5}$, R.~L.~Webster$^{13,2}$, S. B. Wyithe$^{13,2}$, 
R.~J.~Cappallo$^{20}$, A.~A.~Deshpande$^{9}$, D.~L.~Kaplan$^{21}$, C.~J.~Lonsdale$^{20}$, S.~R.~McWhirter$^{20}$, E.~Morgan$^{16}$, D.~Oberoi$^{22}$, S.~M.~Ord$^{5,7,2}$, T.~Prabu$^{9}$, K.~S.~Srivani$^{9}$, A.~Williams$^{5}$,  C.~L.~Williams$^{16}$ 
}
\email{\tt anna.kapinska@uwa.edu.au}

\affil{
$^{1}$ International Centre for Radio Astronomy Research (ICRAR), University of Western Australia, 35 Stirling Hwy, WA 6009, Australia\\
$^{2}$ ARC Centre of Excellence for All-Sky Astrophysics (CAASTRO)\\
$^{3}$ Research School of Astronomy and Astrophysics, Australian National University, Canberra, ACT 2611, Australia\\
$^{4}$ Center of Astrophysics and Supercomputing, Swinburne University of Technology, VIC 3122, Australia\\
$^{5}$ International Centre for Radio Astronomy Research (ICRAR), Curtin University, Bentley, WA 6102, Australia\\
$^{6}$ Netherlands Institute for Radio Astronomy (ASTRON), PO Box 2, 7990 AA Dwingeloo, The Netherlands\\
$^{7}$ CSIRO Astronomy and Space Science (CASS), PO Box 76, Epping, NSW 1710, Australia\\
$^{8}$ Sydney Institute for Astronomy, School of Physics, The University of Sydney, NSW 2006, Australia\\
$^{9}$ Raman Research Institute, Bangalore 560080, India\\
$^{10}$ Dunlap Institute for Astronomy and Astrophysics, University of Toronto, ON, M5S 3H4, Canada\\
$^{11}$ Centre of Astrophysics Research, University of Hertfordshire, College Lane, Hatfield AL10 9AB, UK\\
$^{12}$ School of Chemical \& Physical Sciences, Victoria University of Wellington, PO Box 600, Wellington 6140, New Zealand\\
$^{13}$ School of Physics, The University of Melbourne, Parkville, VIC 3010, Australia\\
$^{14}$ Department of Physics, University of Washington, Seattle, WA 98195, USA\\
$^{15}$ School of Earth and Space Exploration, Arizona State University, Tempe, AZ 85287, USA\\
$^{16}$ MIT Haystack Observatory, Westford, MA 01886, USA\\
$^{17}$ Harvard-Smithsonian Center for Astrophysics, Cambridge, MA 02138, USA\\
$^{18}$ Department of Physics, Brown University, Providence, R1 02906, USA\\
$^{19}$ Instituto di Radio Astronomia, Instituto Nationale di Astrophysica, Bologna, Italy\\
$^{20}$ Kavli Institute for Astrophysics and Space Research, Massachusetts Institute of Technology, Cambridge, MA 02139, USA\\
$^{21}$ Department of Physics, University of Wisconsin--Milwaukee, Milwaukee, WI 53201, USA\\
$^{22}$ National Centre for Radio Astrophysics, Tata Institute for Fundamental Research, Pune 411007, India
}

\begin{abstract}

 We present new radio continuum observations of NGC~253 from the Murchison Widefield Array at frequencies between 76 and 227~MHz. We model the broadband radio spectral energy distribution for the total flux density of NGC~253 between 76~MHz and 11~GHz. The spectrum is best described as a sum of central starburst and extended emission. The central component, corresponding to the inner 500~pc of the starburst region of the galaxy, is best modelled as an internally free-free absorbed synchrotron plasma, with a turnover frequency around 230~MHz. The extended emission component of the NGC~253 spectrum is best described as a synchrotron emission flattening at low radio frequencies. We find that 34\% of the extended emission (outside the central starburst region) at 1~GHz becomes {  partially} absorbed at low radio frequencies. Most of this flattening occurs in the western region of the SE halo, and may be indicative of synchrotron self-absorption of shock re-accelerated electrons or an intrinsic low-energy cut off of the electron distribution. Furthermore,  we detect the large-scale synchrotron radio halo of NGC~253 in our radio images.  At 154--231~MHz the halo displays the well known X-shaped/horn-like structure, and extends out to $\sim 8$~kpc in $z$-direction (from major axis).

\end{abstract}



\section{INTRODUCTION}
\label{sec:intro}

Observing at low radio frequencies ($\lesssim0.5$~GHz) is of a particular value; low energy and old plasma can be revealed, tracing and constraining physical conditions in galaxies. In star forming galaxies the low surface brightness plasma forms e.g. extended halos associated with winds and large scale magnetic fields, or diffuse emission from galactic disks. Furthermore, measurements at low radio frequencies can help to distinguish for instance between thermal and non-thermal plasma, and their absorbing mechanisms, responsible for the level of observed radio emission. It is expected that the Square Kilometre Array (SKA) will unravel a large star-forming galaxy population \cite[e.g.][]{2015aska.confE..70B,2015aska.confE..68J}, but before we can embark on a large scale study of star-forming and starburst galaxies and their evolution with continuum radio surveys, we need to understand the origin of the complex radio spectral energy distributions and morphologies of these galaxies. Nearby objects are ideal laboratories for this task.

NGC~253 is the dominant galaxy in the nearby Sculptor Group, at a distance of 3.94~Mpc from the Local Group \cite[][]{2003AA...404...93K} and velocity $cz=240$~km~s$^{-1}$. It is an almost edge-on SBc type galaxy \cite[][]{1991rc3..book.....D} observed at an inclination of $78.5^{\circ}$ \cite[][]{1980ApJ...239...54P} and is considered a prototype of nuclear starburst galaxies \cite[][]{1980ApJ...238...24R}. Its estimated stellar mass is $\sim4\times10^{10}$~M$_{\odot}$, with a prominent stellar halo of $2.5\times10^{9}$~M$_{\odot}$ extending up to 30~kpc above the disk \cite[][]{2011ApJ...736...24B}.  As one of the closest and most prominent galaxies, NGC~253 has been extensively studied in all wavelengths, including broadband radio continuum, polarization and {  H{\sc i}} observations \cite[][among others]{1979AA....77...25B,1983AA...127..177K,1984A&A...137..138H,1992ApJ...399L..59C,1994A&A...292..409B,1997ApJ...488..621U,2006AJ....132.1333L,2009A&A...494..563H,2009A&A...506.1123H,2011A&A...535A..79H,2015MNRAS.450.3935L}. 

Radio emission from starburst galaxies originates from two principal components: the non-thermal synchrotron emission from relativistic electrons spiralling in the interstellar magnetic field, and the thermal emission from electrons colliding with ions in the ionized interstellar medium (ISM) around hot stars. The sources of the non-thermal emission are predominantly cosmic rays accelerated by supernova remnants (SNR) that in NGC~253 ultimately create a prominent synchrotron radio halo \cite[][]{1992ApJ...399L..59C}. {  Studies of the NGC~253 magnetic field suggest the disk wind model and large-scale dynamo action to be shaping the vertical structure of the field, which in turn enhances the cosmic ray transport through a collimation of strong, starburst driven superwind} \cite[][]{1994A&A...292..409B,2000A&A...360...24P,2002ApJ...568..689S,2009A&A...494..563H,2009A&A...506.1123H,2011A&A...535A..79H}.

The starburst region of NGC~253 is violently active; the supernova rate of the inner 300~pc of the galaxy is estimated to be between $0.14$ and $2.4$~yr$^{-1}$, and the star formation rate is $\sim5$~M$_{\odot}$~yr$^{-1}$ \cite[][]{2006AJ....132.1333L,2014AJ....147....5R,2015MNRAS.450L..80B,2015MNRAS.450.3935L}. It has been suggested that up to half of the radio sources in the central starburst region are dominated by thermal emission: i.e. H{\sc ii} regions characterized by a flat radio spectral index\footnote{Radio spectral index $\alpha$ is defined such that the flux density $S_\nu$ at a frequency $\nu$ is $S_\nu \propto \nu^{-\alpha}$.} $\alpha \simeq 0.1$ and including at least one large supercluster of stars \cite[][]{1997ApJ...488..621U,1999ApJ...518..183K}. Outside the central starburst region the radio emission at GHz frequencies is dominated by steep spectrum diffuse emission and SNRs, but several strong thermal sources are detected \cite[][]{2000AJ....120..278U}. Based on integrated radio continuum spectra, \cite{1997A&A...322...19N} estimated 10\% of the NGC~253 flux density at 1~GHz to be of thermal origin, increasing to 35\% at 10~GHz. 

At low radio frequencies both of these principal components become pronounced. Synchrotron emission has steep spectrum, becoming dominant at sub-GHz frequencies due to the population of old, low energy electrons. However, such emission may be also subject to self-absorption in the case of compact objects. Thermal emission also becomes increasingly more absorbed with decreasing frequency. The free-free absorption in the central starburst of NGC~253 has previously been measured \cite[][]{1996A&A...305..402C,2004AJ....127...10T}.

Here, we present extensive low radio frequency ($<230$~MHz) imaging of NGC~253 obtained with the Murchison Widefield Array \cite[MWA;][]{2013PASA...30...31B,2013PASA...30....7T}. Our images are some of the deepest yet at these frequencies, and at low angular resolution they are especially sensitive to large-scale diffuse structure, allowing us to investigate the extent and frequency dependence of the radio halo. 
The paper is structured as follows. Our radio data and methods, including assumed models of radio spectra and model fitting, are described in \S\ref{sec:data} and \S\ref{sec:methods} respectively. Results are presented in \S\ref{sec:results}. The synchrotron radio halo of NGC~253 is discussed in \S\ref{sec:halo}. We discuss low frequency radio emission from NGC~253, its radio spectral energy distribution and radio spectral maps in \S\ref{sec:spectral-props}. Conclusions are given in \S\ref{sec:conclude}.


\section{Observations and data reduction}
\label{sec:data}

We use radio continuum data from the Galactic and Extragalactic All-Sky MWA Survey \cite[GLEAM;][]{2015PASA...32...25W} and the MWA Epoch of Reionization experiment \cite[MWA/EoR;][]{2013PASA...30...31B,2016ApJ...819....8P}. The GLEAM survey provides unprecedented spectral coverage between 72 and 231~MHz, while the MWA/EoR image at 169~MHz is {  almost twice as deep as the most sensitive GLEAM image at 200~MHz (rms noise 4.1~mJy beam$^{-1}$ and 7.3~mJy beam$^{-1}$ respectively)}. In addition, the data have been observed and processed independently, providing a verification of our flux density calibration.

\subsection{The GaLactic and Extragalactic All-Sky MWA Survey (GLEAM)}

The GLEAM survey observed the entire radio sky south of declination $+30^{\circ}$ at an angular resolution of approximately 1.7~arcmin (227~MHz) to 5~arcmin (76~MHz).  At 154~MHz the GLEAM survey is sensitive to structures up to $10$~deg in angular scale, and has an instantaneous field of view of $25\times25$~deg$^{2}$. The observations were made in a meridian drift scan mode covering frequencies between 72 and 231 MHz with bandwidths of 7.68~MHz grouped in five 30.72~MHz-wide bands. These bands, centred on 87.7, 118.4, 154.2, 185.0 and 215.7~MHz (hereafter 88, 118, 154, 185 and 216~MHz), were observed sequentially as 112 sec snapshots; each frequency was observed every 10~min. During a night typically 8--10 h in hour angle were observed. Frequencies between 134 and 137~MHz were avoided due to satellite interference. For more details on the survey parameters and strategy see \cite{2015PASA...32...25W}. 

Here we use GLEAM data from the first year of observing \cite[Data Release 1 from 2013 August -- 2014 June;][]{2016MNRAS.GLEAM.subm}. The sky area covering NGC~253 was observed on 2013 August 10 and 2013 November 25. The full data reduction process is described in detail in \cite{2016MNRAS.GLEAM.subm}; here we summarize only the main calibration and imaging steps. 

The correlated data were first pre-processed with the {\sc cotter} pipeline which performs flagging of data affected by radio frequency interference (RFI) and averaging of the data to 1s time and 40~kHz frequency resolution  \cite[][]{2010MNRAS.405..155O,2015PASA...32....8O}. Standard calibration (phase and amplitude bandpass calibration) was done with {\sc CASA} \cite[Common Astronomy Software Applications package; ][]{2007ASPC..376..127M}. Imaging and self-calibration were then performed using {\sc WSClean} imager \cite[][]{2014MNRAS.444..606O} that corrects for wide field $w$-term effects. Images of a 7.68~MHz bandwidth at 20 frequencies continuously distributed between 72 and 231~MHz (avoiding 134--137~MHz) and using a robust weighting $r=-1.0$ \cite[][]{1995AAS...18711202B} were then  created. Deconvolution has been performed at this stage, and details are provided in \cite{2016MNRAS.GLEAM.subm}.

The primary beam correction of our GLEAM observations was done with the \cite{2015RaSc...50...52S} model down to the 10\% level of the beam response. An additional calibration stage was necessary to correct for residual declination dependence of the flux density scale in the final mosaics arising from the limited accuracy of the adopted primary beam model. This was done by comparing flux density measurements of all unresolved sources extracted from GLEAM images above $8\sigma$ rms noise level to their radio spectra as predicted by three catalogues: VLA Low-Frequency Sky Survey redux \cite[VLSSr;][]{2014MNRAS.440..327L}, MRC and NRAO VLA Sky Survey  \cite[NVSS;][]{1998AJ....115.1693C}.  The absolute flux density scale of the GLEAM images is accurate to 8\%, which is included in the quoted uncertainties of the measurements \cite[for details see][]{2016MNRAS.GLEAM.subm}.

Images for each of five central frequencies centered on frequencies of 88, 118, 154, 185 and 216~MHz and of bandwidth 30.72~MHz were made. The two highest frequency images are further combined to create a `deep' image at 200~MHz with a 61.4~MHz bandwidth. We also use the 7.68-MHz images for construction of the high resolution radio spectrum of NGC~253. The  final synthesised beam sizes, rms and background noise levels in the deep 200~MHz image are $2.22\times2.12$~arcmin$^2$, ${\rm PA}=-78^\circ$, 11~mJy~beam$^{-1}$ and 7.3~mJy~beam$^{-1}$ respectively, and {  their range between the lowest and highest GLEAM frequencies is listed in Table~\ref{tab:noisebm}}.


\begin{table}
  \scriptsize
  \centering
  \caption{  Range of angular resolution and noise values of the GLEAM data. }
  \label{tab:noisebm}
\begin{tabular}{ccccccc}
\hline
\hline
 $\nu$  & \multicolumn{2}{c}{synthesised beam}  & rms noise & background \\
 (MHz)  &  bmaj$\times$bmin   & PA              & (mJy beam$^{-1}$) & noise\\
        &  (arcmin$^2$)       & (deg)           &                  & (mJy beam$^{-1}$)\\
\hline
     76 &  $5.03\times4.72$   & $-18.8$  &  $107$  &  $-44$ \\
    227 &  $1.73\times1.67$   & $-26.0$  &  $12.8$ &  $-3.3$ \\
\hline\\
\end{tabular}
\end{table}


\begin{figure}
\includegraphics[width=86mm]{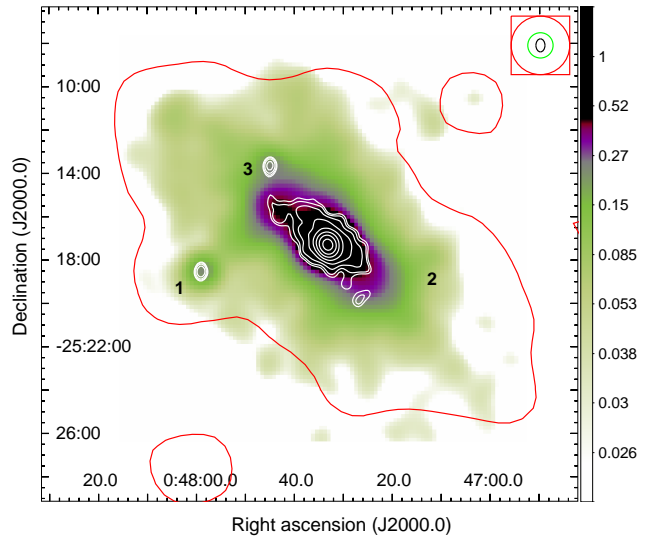}
\caption{The 330~MHz image of NGC~253 from \cite{1996A&A...305..402C} with overlaid contours from the TGSS ADR1 survey (white) and the MWA/EoR image (red). The TGSS contours start at $4\sigma$ local rms noise level ($\sigma=11.7$~mJy~beam$^{-1}$) and increase as $\sigma2^i$ for $i>0$. The MWA/EoR0  contour marks the $4\sigma$ radio intensity at 169~MHz (16.4~mJy~beam$^{-1}$). The sizes of the synthesised beams at 169~MHz (red), 330~MHz (green) and 150~MHz (black) are drawn in the top right corner. Background sources, not associated with the intrinsic emission of NGC~253, are labelled with numbers (see \S\ref{sec:bkg-srcs}). The color scale is in units of Jy beam$^{-1}$, and the pixel size is $5\times5$~arcsec$^2$.\\}
\label{rys:tgss-carilli}
\label{rys:carilli-halo}
\label{rys:bkg-sources}
\end{figure}


\subsection{MWA Epoch of Reionization (EoR) data}

The observed MWA/EoR field that contains the Sculptor Group (EoR0 field) is centered on $\text{RA} = 0^{\rm h}$, $\text{Dec} = -27^{\circ}$, and was observed for a total of 30 hours between August and October 2013 in a combination of a tracking and drifting modes. In this hybrid mode the telescope tracks a set of discrete pointing centers through which the field of interest is drifting. The observations cover frequencies between $138.9-197.7$~MHz observed as two bands (low and high) each with an instantaneous bandwidth of $\Delta\nu=30.72$~MHz.

The correlated MWA/EoR0 data were pre-processed with the {\sc cotter} pipeline \cite[][]{2015PASA...32....8O} and averaged to 4s time and 40~kHz frequency resolution. Calibration of the data was performed as a direction-independent self-calibration using the {\sc mitchcal} tool \cite[][]{2008ISTSP...2..707M} and was based on a bootstrapped sky model. The initial sky model was generated from the MWA Commissioning Survey \cite[][]{2014PASA...31...45H}, the MRC catalogue and the Sydney University Molonglo Sky Survey \cite[SUMSS;][]{2003MNRAS.342.1117M}. Imaging was performed with the {\sc WSClean} software that corrects for the non-zero $w$-term effects. During the imaging process 2,500 sources were peeled, and the images were created with a uniform weighting. The primary beam, and so the flux density scale, was corrected by applying the \cite{2015RaSc...50...52S} model. As shown by \cite{2014PASA...31...45H}, this model is accurate to 10\%, hence we add this error in quadrature to the quoted uncertainties of our measurements.

The final image used in this paper is centred at 169.6~MHz (thereafter 169~MHz) with a total bandwidth of $\Delta\nu=58.8$~MHz, synthesised beam size $2.3\times2.3$~arcmin$^2$ and rms noise $4.1$~mJy~beam$^{-1}$. The calibration and imaging process of the EoR0 data is presented and discussed in detail in \cite{2016MNRAS.458.1057O}.

\subsection{Other low frequency radio surveys}

There are additional two all-sky low frequency radio surveys that include NGC~253: the 74~MHz VLSSr \cite[][]{2014MNRAS.440..327L} and the 150~MHz Tata Institute of Fundamental Research (TIFR) Giant Metrewave Radio Telescope (GMRT) Sky Survey {  (TGSS) Alternative Data Release~1} \cite[ADR1;][]{2016arXiv160304368I}.

The TGSS survey observed the whole radio sky north of declination $-53^{\circ}$ at a frequency 150~MHz (bandwidth $\Delta\nu=16.7$~MHz) at an angular resolution $25\times 25/\text{cos}(\delta-19^{\circ})$~arcsec$^2$ at declinations south of $+19^{\circ}$. The instantenous field of view of the survey at half power at 150~MHz is $3.1\times3.1$~deg$^2$, with sensitivity to structures up to 68~arcmin in angular scale \cite[][]{2016arXiv160304368I}. Since the absolute flux density calibration of the TGSS ADR1 may be uncertain up to 50\% in some sky regions\footnote{\tt http://tgssadr.strw.leidenuniv.nl/doku.php}, we independently verified the calibration in the area of NGC~253. We selected unresolved sources with flux density $>1$~Jy from the $5\times5$~deg$^2$ mosaic that included NGC~253 (R03\_D17). We compared the TGSS ADR1 flux densities of these sources with the predicted values based on the spectral modeling in which we used the VLSSr, GLEAM (deep 200~MHz), MRC and NVSS surveys. We found that the TGSS mosaic required scaling by a factor 1.02 in flux density, and the absolute flux density calibration was accurate to 7\%; we further added this error in quadrature to the quoted uncertainties of our measurements. 

The VLSS survey \cite[][]{2007AJ....134.1245C} observed the radio sky north of $-30^{\circ}$ at a frequency 74~MHz. Here we use the recent re-reduction of the survey data, the VLSSr \cite[][]{2014MNRAS.440..327L}. VLSSr images have an angular resolution of $75\times75$~arcsec$^2$ and a theoretical sensitivity to structures of 13--37~arcmin in angular scale. 

We find that neither TGSS nor VLSSr are sensitive to the extended emission of NGC~253 (Figure~\ref{rys:tgss-carilli}). For this reason we use the TGSS data for the flux density measurement of the central starburst region, and both TGSS and VLSSr for measurements of the flux densities of background sources only.


\begin{figure*}
\includegraphics[width=180mm]{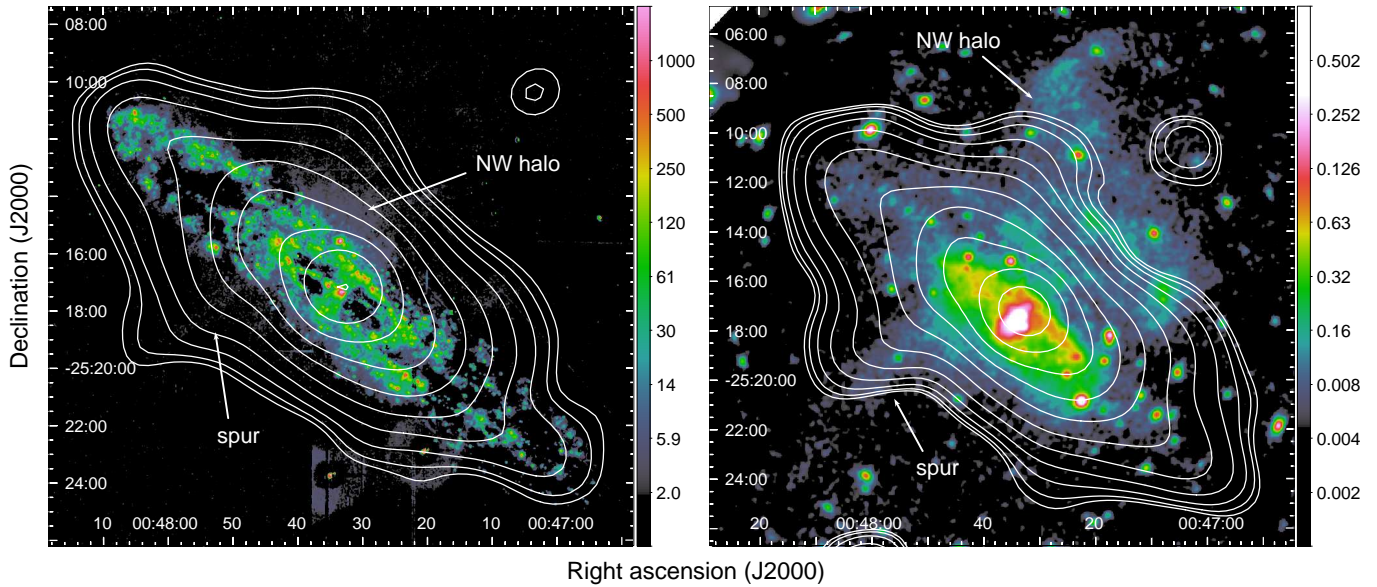}
\caption{{\it Left:} H$\alpha$ image from the Survey for Ionization in Neutral Gas Galaxies \protect\cite[SINGG;][]{2006ApJS..165..307M}. The image was taken with the CTIO 0.9m telescope for SINGG using a 75\AA-wide narrow band filter, and continuum was subtracted using an image taken in the $R$ band (Harris filter) in the same observing setup. The image has been smoothed by a factor {  6} to aid in inspection of the faint wind features. The GLEAM contours start at $4\sigma$ rms level ($\sigma=11$~mJy~bm$^{-1}$) and increase by a factor of $\sigma 2^i$ for $i>0$. The background discrete sources within extended emission of NGC~253 have been removed (see \S\ref{sec:bkg-srcs}). {\it Right:} {\it XMM-Newton} image of soft X-ray emission from NGC~253 and its environment \cite[0.2--1 keV band; ][]{2008AA...489.1029B} with overlaid MWA/EoR0 intensity contours. {  Both color bars} are in units of $10^{-3}$~ct~s$^{-1}$~px$^{-1}$. The MWA/EoR0 contours start at $4\sigma$ rms level ($\sigma=4.1$~mJy/bm) and increase by a factor of $\sigma 2^i$ for $i>0$.
\\}
\label{rys:multi-images}
\end{figure*}

\section{Methods}
\label{sec:methods}

\subsection{Flux density measurements}
\label{sec:flux-dens}

Measurements of the total flux density of NGC~253 were performed with {\sc CASA} task {\sc imstat} that provides a summed flux density within a specified regions of the image corrected for the synthesised beam. We masked all pixels below $2.6\sigma$ local rms noise level {  \cite[][]{2012MNRAS.425..979H}}. For point sources the flux density was measured with the {\sc AIPS} task {\sc jmfit}; for each unresolved source we fit for two components, a Gaussian and a zero-level with a slope. The absolute flux density scale is set to the \cite{1977AA....61...99B} scale.

\subsection{Models of radio spectra}
\label{sec:models-spectra}

Radio sources often show simple spectra that can be approximated by a power-law. However, at low radio frequencies (a few hundred MHz and below) radio spectra are consistently more curved \cite[][]{1980MNRAS.190..903L,1991A&A...251..442H,1993ApJ...410..626D,1999AJ....117..677B,2012MNRAS.421..108D,2015AJ....149...32M} until a turnover frequency below which the spectra become inverted. The spectral turnover is typically caused by either synchrotron self-absorption and/or thermal free-free absorption \cite[][and references therein]{2015ApJ...809..168C}. If there is no evidence for a spectral turnover in the radio spectra we construct here { (see \S\ref{sec:modselct} for model selection method)}, we proceed with fitting a polynomial. The curved radio spectra are then modeled with an $n$th-order polynomial, that in the logarithmic scale takes a general form of 
\begin{equation}
  \begin{aligned}
    \text{log}(S_{\nu}) & = \sum^n_{i=0} A_i \text{log}^i\left(\frac{\nu}{\nu_0}\right) \\
                       & = A_0 + A_1 \text{log}\left(\frac{\nu}{\nu_0}\right) + ... + A_n \text{log}^n\left(\frac{\nu}{\nu_0}\right),
  \end{aligned}
\label{eqn:poly}
\end{equation}
where $A_0$ is an offset parameter (equivalent to log($S_0$) in the simple power-law case), $A_1$ is the spectral index $-\alpha$, and $A_n$ are curvature parameters ($c_n$). In the linear space the model takes the following form
\begin{equation}
S_\nu = \prod^n_{i=0} 10^{A_i\text{log}^i(\nu/\nu_0)},
\end{equation}
which we use in our modeling to preserve Gaussian noise characteristics of the measurements.

Where the data suggest or show a spectral turnover, the following models are tested: synchrotron self-absorption, free-free absorption or a combination of these and power-law components.


\subsubsection{Synchrotron self-absorption (SSA)}

At low radio frequencies the intensity of the synchrotron radiation may become sufficiently high (optically thick regime) for re-absorption, termed synchrotron self-absorption, to take over. The process may be important, or even dominant, for compact sources \cite[][]{1990SvAL...16..339S,1998ApJ...499..810C}. We model the synchrotron radio spectra that may turnover due to self-absorption at low radio frequencies as \cite[e.g.][]{2003AJ....126..723T}
\begin{equation}
S_\nu = S_{\tau=1} \left( \frac{\nu}{\nu_{\tau=1}} \right)^{-\alpha} \left ( \frac{1 - e^{-\tau(\nu)}}{\tau(\nu)}\right ),
\label{eqn:SSA}
\end{equation}
\begin{equation}
\tau(\nu) = (\nu/\nu_{\tau=1})^{-(\alpha+2.5)},
\end{equation}
where $\nu_{\tau=1}$ is a frequency at which the optical depth ($\tau$) reaches unity.


\begin{table*}
  \scriptsize
  \centering
  \caption{  Flux density measurements of background sources embedded in NGC~253 radio emission. All measurements are in the same absolute flux density scale of \cite{1977AA....61...99B}. \vspace{-4mm}}
\label{tab:spectra-bkg}
\begin{center}
\begin{tabular}{lrrrcc}
\hline
\hline
Frequency & \multicolumn{1}{c}{background} & \multicolumn{1}{c}{background} & \multicolumn{1}{c}{background} & {Angular}    & References\\
$[$MHz$]$ & \multicolumn{1}{c}{source 1}   & \multicolumn{1}{c}{source 2}   & \multicolumn{1}{c}{source 3}   & {resolution} & \\
          & \multicolumn{1}{c}{(mJy)}      & \multicolumn{1}{c}{(mJy)}      & \multicolumn{1}{c}{(mJy)}      & (arcsec$^2$) &  \\
\hline
74    & $263 \pm76^{\dagger}$& &                     & $75\times75$  & a, \textreferencemark\\
150   & $212 \pm19$  & $15.6^{\rm up}$ & $162\pm17$  & $36\times24$  & b, \textreferencemark \\
200   & $233 \pm23$  & &                           & $138\times126$& c\\
330   & $190\pm10$   & &                           & $72\times72$  & d \\
610   & $84 \pm15$   & &                           & $114\times24$ & e \\
843   & $97 \pm10$   & &                           & $47\times43$  & f \\
1465  & $63.0\pm2.5$ & &                           & $66\times38$  & g \\
1465  & $53.0\pm3.0$ & $17.2\pm1.0$  & $26\pm2$    & $30\times30$  & d \\
1490  & $5.6 \pm0.5$ &               & $53.0\pm0.5$& $19.0\pm0.5$  & h \\
4850  & $19.0\pm1.0$ & $6.6\pm0.3$   & $7.9\pm0.4$ & $30\times30$  & d \\
8350  & $9.5\pm0.5$  & &                           & $84\times84$  & d \\
\hline
\end{tabular}
\end{center}
\vspace{-2mm}
$\dagger$ Tentative detection ($2.5\sigma$). $^{\rm up}$ Upper limit, equal $3\times$ local rms noise level. \textreferencemark~Our measurement based on images from the quoted survey. 
{\bf  References.} 
          (a) \protect\cite{2014MNRAS.440..327L}, 
          (b) \protect\cite{2016arXiv160304368I}, 
          (c) This publication, 
          (d) \protect\cite{2009A&A...494..563H},
          (e) \protect\cite{1983A&AS...54..387B},
          (f) \protect\cite{1983PASAu...5..235R},
          (g) \protect\cite{1984A&A...137..138H},
          (h) \protect\cite{1987ApJS...65..485C}.
\end{table*}

\begin{table*}
  \scriptsize
  \centering
  \caption{Results of the spectral modeling of the background sources. }
  \label{tab:bkgmodels}
\begin{tabular}{ccccccc}
\hline
\hline
Background & polynomial & $S_{1 \rm GHz}$ & $\alpha$ & $c_1$ & dof & $\chi^2$ \\ 
source    & order      & (mJy)         &          &       &     & \\
\hline
1  & 2 & $80.7 \pm1.9$  &  $0.79 \pm0.02$ &  $-0.25 \pm0.04$ & 7 & 19.2 \\
2  & 2 & $21.9 \pm1.0$  &  $0.32 \pm0.07$ &  $-0.63 \pm0.11$ & 1 &  3.5\\
3  & 1 & $33.1 \pm1.4$  &  $0.88 \pm0.03$ &  --              & 1 &  2.8\\
\hline
\end{tabular}
\end{table*}


\begin{table*}
\caption{{  Total integrated flux density measurements of NGC~253.} Uncertainties associated with the GLEAM, MWA/EoR and TGSS  measurements include the fitting error and the uncertainty of the absolute flux density calibration, for the remaining data the errors quoted in references are adopted. All measurements are brought to the same absolute flux density scale of \cite{1977AA....61...99B}. {  If no information on the largest angular scale of the final images was reported, we state `no info', and provide a theoretical value (upper limit, marked as `th') for those measurements for which information on interferometer's shortest baseline was provided, and assuming that adequate $uv$ coverage has been achieved and no minimum-$uv$ cut has been applied.} \vspace{-3mm}}
\label{tab:spectra}
\begin{center}
\begin{tabular}{lrccc}
\hline
\hline
Frequency & \multicolumn{1}{c}{total} & {  Angular}     & {  Largest      } & References\\
$[$MHz$]$ & \multicolumn{1}{c}{(Jy)}  & {  resolution}  & {  angular scale} &\\
          &                           & (arcmin$^2$)      & (deg)               & \\
\hline
76  & $23.9\pm2.0$ & $5.1\times4.7$ & 29      & a\\
80  & $23.7\pm4.0$ & $3.7\times3.7$ & no info & b,c \\ 
84  & $22.6\pm1.9$ & $5.1\times4.7$ & 27      & a\\
92  & $20.9\pm1.7$ & $5.1\times4.7$ & 24      & a\\
99  & $20.0\pm1.6$ & $5.1\times4.7$ & 23      & a\\
107 & $20.3\pm1.7$ & $5.1\times4.7$ & 21      & a\\
115 & $20.0\pm1.6$ & $5.1\times4.7$ & 19      & a\\ 
122 & $18.8\pm1.6$ & $5.1\times4.7$ & 18      & a\\
130 & $17.8\pm1.5$ & $5.1\times4.7$ & 17      & a\\
143 & $17.9\pm1.5$ & $5.1\times4.7$ & 16      & a\\
151 & $16.6\pm1.4$ & $5.1\times4.7$ & 15      & a\\
158 & $16.1\pm1.3$ & $5.1\times4.7$ & 14      & a\\
166 & $15.7\pm1.3$ & $5.1\times4.7$ & 13      & a\\ 
169 & $15.0\pm1.2$ & $5.1\times4.7$ & 13      & a\\
174 & $16.4\pm1.3$ & $5.1\times4.7$ & 13      & a\\
181 & $15.7\pm1.3$ & $5.1\times4.7$ & 12      & a\\
189 & $15.3\pm1.2$ & $5.1\times4.7$ & 12      & a\\
197 & $15.4\pm1.2$ & $5.1\times4.7$ & 11      & a\\
204 & $15.8\pm1.3$ & $5.1\times4.7$ & 11      & a\\
212 & $15.1\pm1.2$ & $5.1\times4.7$ & 11      & a\\
220 & $14.7\pm1.2$ & $5.1\times4.7$ & 10      & a\\
227 & $15.0\pm1.2$ & $5.1\times4.7$ & 10      & a\\
330   & $16.5 \pm1.9$  & $1.2\times 1.2$   & 1.2                   & d, e \\
408   & $15.7 \pm1.9$  & $2.9\times2.86$   & no info               & f \\
468   & $15.1 \pm1.5$  & $2.1\times2.1$, $5.2\times5.2 ^\diamondsuit$ & extr. single dish$^\clubsuit$ & c, g \\ 
610   &  $9.4 \pm0.6$  & $1.9\times0.4$    & no info               & h \\
843   &  $9.0 \pm0.9$  & $0.72\times0.78$  & no info (th: 1.1)     & i \\
960   &  $8.0 \pm0.12$ & $20.2\times20.2$  & single dish           & c, g \\ 

1100  &  $6.7 \pm0.08$ & $4.2\times4.2$    & no info, dense core   & j \\ 
1200  &  $6.68 \pm0.10$& $3.4\times3.4$    & no info, dense core   & j \\ 
1300  &  $6.22 \pm0.07$& $3.2\times3.2$    & no info, dense core   & j \\ 
1400  &  $5.89 \pm0.16$& $2.9\times2.9$    & no info, dense core   & j \\ 
1410  &  $6.12 \pm0.12$& $15.5\times15.5$  & single dish           & c, g \\ 
1430  &  $5.7 \pm0.5$  & $0.91\times0.83$  & no info (th: 0.8)     & h \\
1465  &  $5.9 \pm0.1$  & $1.1\times0.63$   & extr. single dish$^\clubsuit$ & k \\
1465  &  $6.3 \pm1.1$  & $0.5\times0.5$    & 0.25                  & d, e\\
1490  &  $5.6 \pm0.5$  & $0.9\times0.9$    & 0.27                  & m \\
2650  & $3.85 \pm0.12$ & $8.3\times8.3$    & single dish           & c, g \\ 
2695  & $4.26 \pm0.14$ & $4.9\times4.9$    & single dish           & n \\
2700  & $3.49 \pm0.12$ & $8.0\times8.0$    & single dish           & c, g \\ 
4850  & $2.93 \pm0.13$ & $2.7\times2.7$    & single dish           & n \\
4850  & $2.71 \pm0.14$ & $0.5\times0.5$    & single dish           & e \\
4850  & $2.69 \pm0.10$ & $4.2\times4.2$    & single dish           & p \\ 
5009  & $2.50 \pm0.23$ & $4.0\times4.0$    & single dish           & c, g \\ 
5009  & $2.12 \pm0.09$ & $4.0\times4.0$    & single dish           & c, r \\ 
8350  & $1.66 \pm0.08$ & $1.4\times1.4$    & single dish           & e \\
8700  & $2.06 \pm0.12$ & $1.5\times1.5$    & single dish           & n \\
10550 & $1.98 \pm0.18$ & $1.2\times1.2$    & single dish           & s \\
10700 & $1.95 \pm0.15$ & $1.2\times1.2$    & single dish           & t \\
\hline
\end{tabular}
\end{center}
\vspace{-2mm}
$\diamondsuit$ Conflicting details given in the reference. $\clubsuit$ Corrected for zero-spacing missing flux density with extrapolation.
{\bf References.} 
          (a) This publication, 
          (b) \protect\cite{1973AuJPA..27....1S},
          (c) \protect\cite{1981AAS...45..367K},
          (d) \protect\cite{1992ApJ...399L..59C},
          (e) \protect\cite{2009A&A...494..563H}, 
          (f) \protect\cite{1971MNRAS.152..403C}, 
          (g) \protect\cite{1975AuJPA..38....1W}, 
          (h) \protect\cite{1983A&AS...54..387B},
          (i) \protect\cite{1983PASAu...5..235R},
          (j) \protect\cite{2010ApJ...710.1462W},
          (k) \protect\cite{1984A&A...137..138H},
          (m) \protect\cite{1987ApJS...65..485C},
          (n) \protect\cite{1979AA....77...25B},
          (p) \protect\cite{1994ApJS...90..179G},
          (r) \protect\cite{1976AuJPA..39....1W},
          (s) \protect\cite{1994A&A...292..409B},
          (t) \protect\cite{1983AA...127..177K}.
\end{table*}


\begin{table}
\caption{  Flux density measurements of NGC~253 central starburst region. All measurements are in the same absolute flux density scale of \cite{1977AA....61...99B}. \vspace{-4mm}}
\label{tab:spectra-core}
\begin{center}
\begin{tabular}{lrcc}
\hline
\hline
Frequency & \multicolumn{1}{c}{central}   & {Angular}    & References\\
$[$MHz$]$ & \multicolumn{1}{c}{component} & {resolution} &\\
          & \multicolumn{1}{c}{(Jy)}      & (arcsec$^2$) & \\
\hline
150   & $2.16\pm0.15$  & $36\times24$  & a, \textreferencemark \\
330   & $2.67 \pm0.16$ & $33\times21$  & b \\
610   & $2.3 \pm0.2$   & $114\times24$ & c \\
1413  & $2.33 \pm0.14\dagger$ & $3\times1.8$  & d \\
1450  & $2.07 \pm0.04$ & $33\times21$  & b \\ 
1465  & $2.04 \pm0.10$ & $30\times30$  & e \\
1660  & $1.96 \pm0.04$ & $33\times21$  & b \\ 
4520  & $1.36 \pm0.04$ & $33\times21$  & b \\ 
4850  & $1.27 \pm0.06$ & $30\times30$  & e \\
4890  & $1.30 \pm0.04$ & $33\times21$  & b \\ 
6700  & $1.13 \pm0.04$ & $37\times37$  & f \\
7000  & $1.04 \pm0.04$ & $35\times35$  & f \\
8090  & $0.93 \pm0.03$ & $33\times21$  & b \\ 
8350  & $0.98 \pm0.05$ & $84\times84$  & e \\
8470  & $0.89 \pm0.03$ & $33\times21$  & b \\ 
\hline
\end{tabular}
\end{center}
$\dagger$ Integrated. \textreferencemark~Our measurement based on images from the quoted survey. 
{\bf References.} 
          (a) \protect\cite{2016arXiv160304368I}, 
          (b) \protect\cite{1996A&A...305..402C},
          (c) \protect\cite{1983A&AS...54..387B},
          (d) \protect\cite{1982ApJ...252..102C},
          (e) \protect\cite{2009A&A...494..563H}, 
          (f) \protect\cite{2010ApJ...710.1462W}. \vspace{2mm}
\end{table}


\begin{figure*}
\begin{center}
\includegraphics[width=180mm]{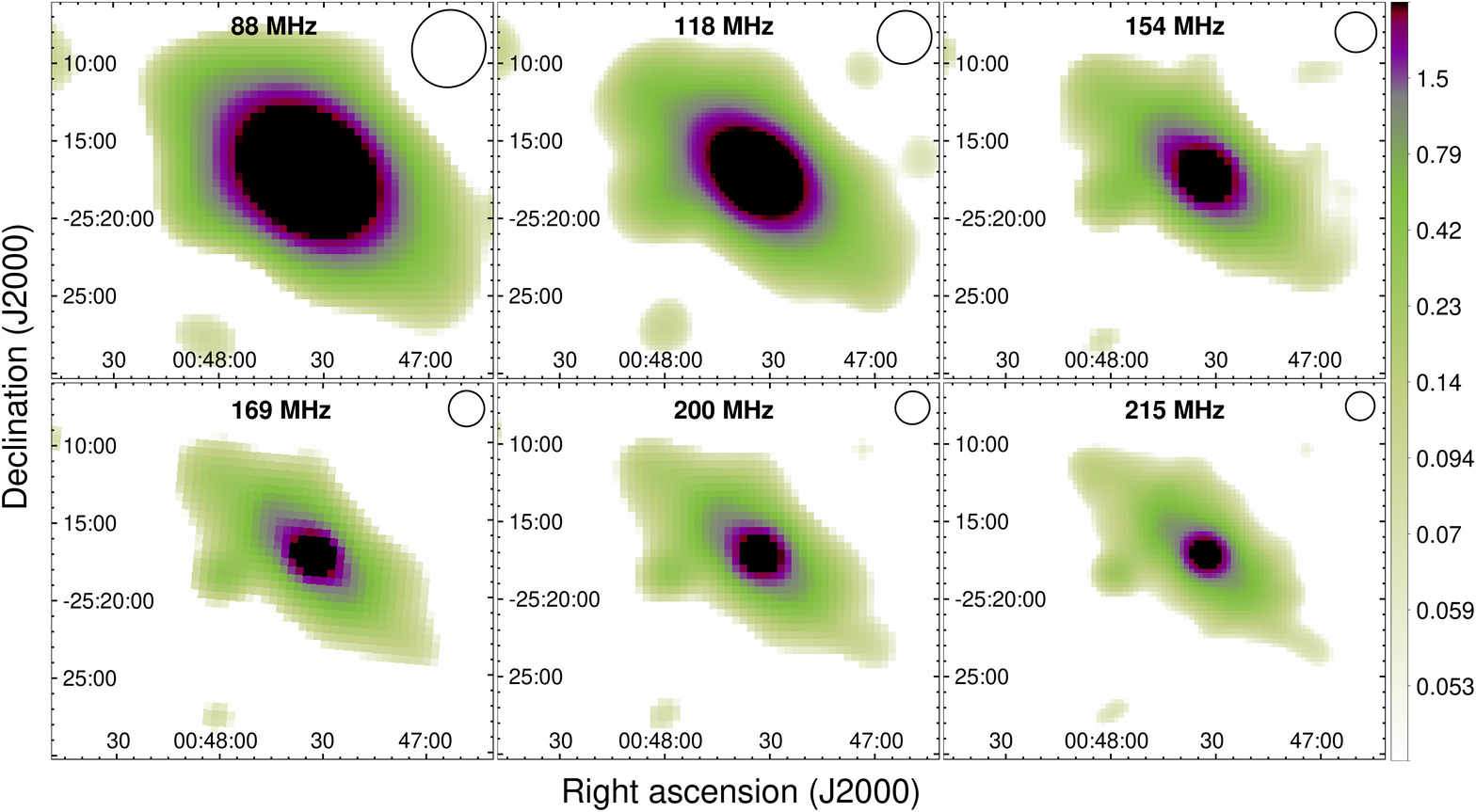}
\caption{Radio images of NGC~253 from the GLEAM survey at five selected frequencies [centered on 88~MHz ({\it top left}), 118~MHz ({\it top middle}), 154~MHz ({\it top right}) and 215~MHz ({\it bottom right}) with bandwidth $\Delta\nu=30.7$~MHz; centered on 200~MHz with bandwidth $\Delta\nu=61.4$~MHz ({\it bottom middle})] and from the MWA EoR data (169~MHz; {\it bottom left}). The synthesised beam sizes are drawn in the top right corner of each image. The color scales are in the units of Jy~beam$^{-1}$, and the pixel sizes are $28.8\times28.8$~arcsec$^2$ at 88~MHz, 118~MHz and 154~MHz, $30.6\times30.6$~arcsec$^2$ at 169~MHz, and $18.8\times18.8$~arcsec$^2$ at 200~MHz and 215~MHz.\\}
\label{rys:radio-images}
\includegraphics[angle=270,width=103mm]{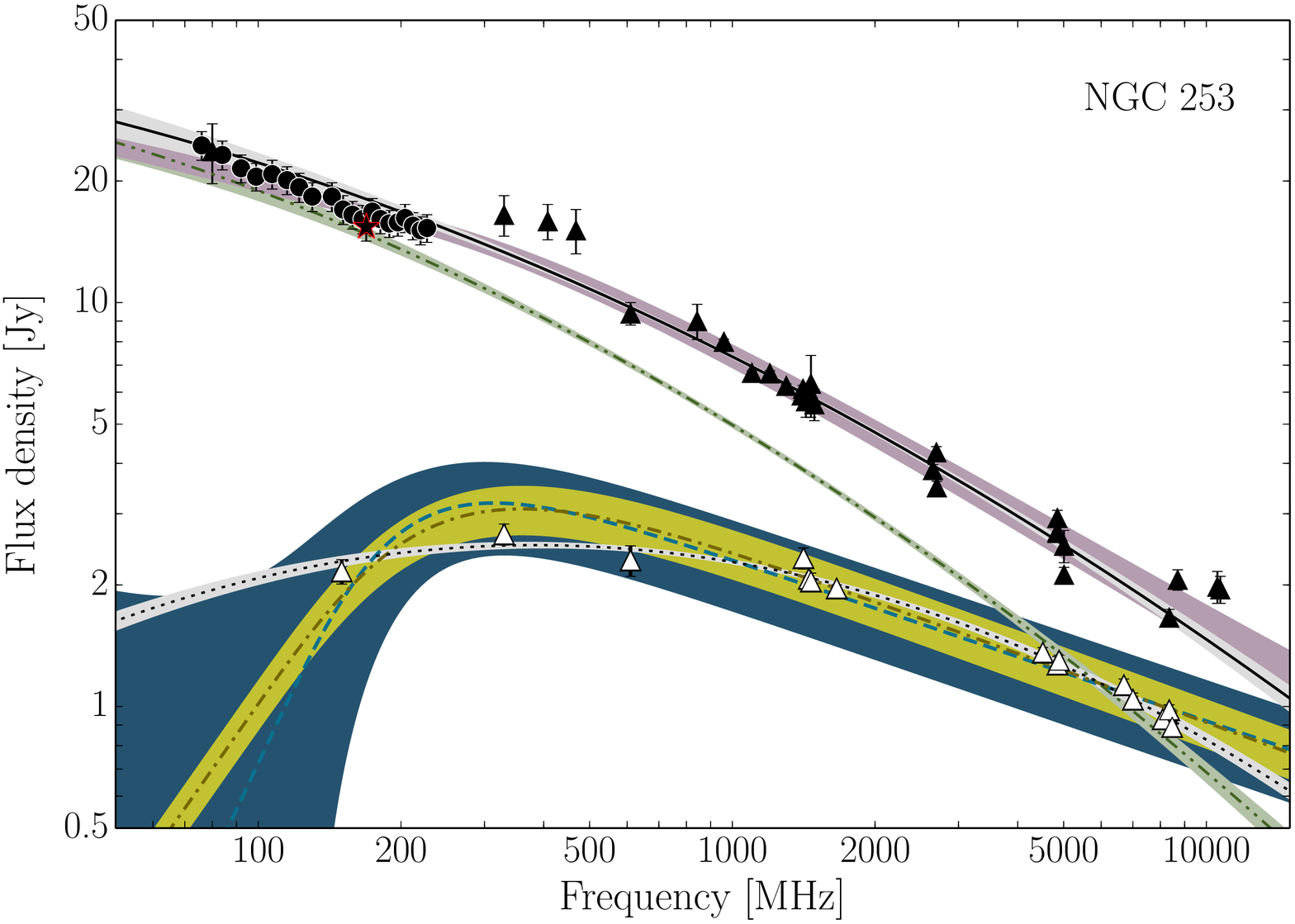}
\includegraphics[angle=270,width=75mm]{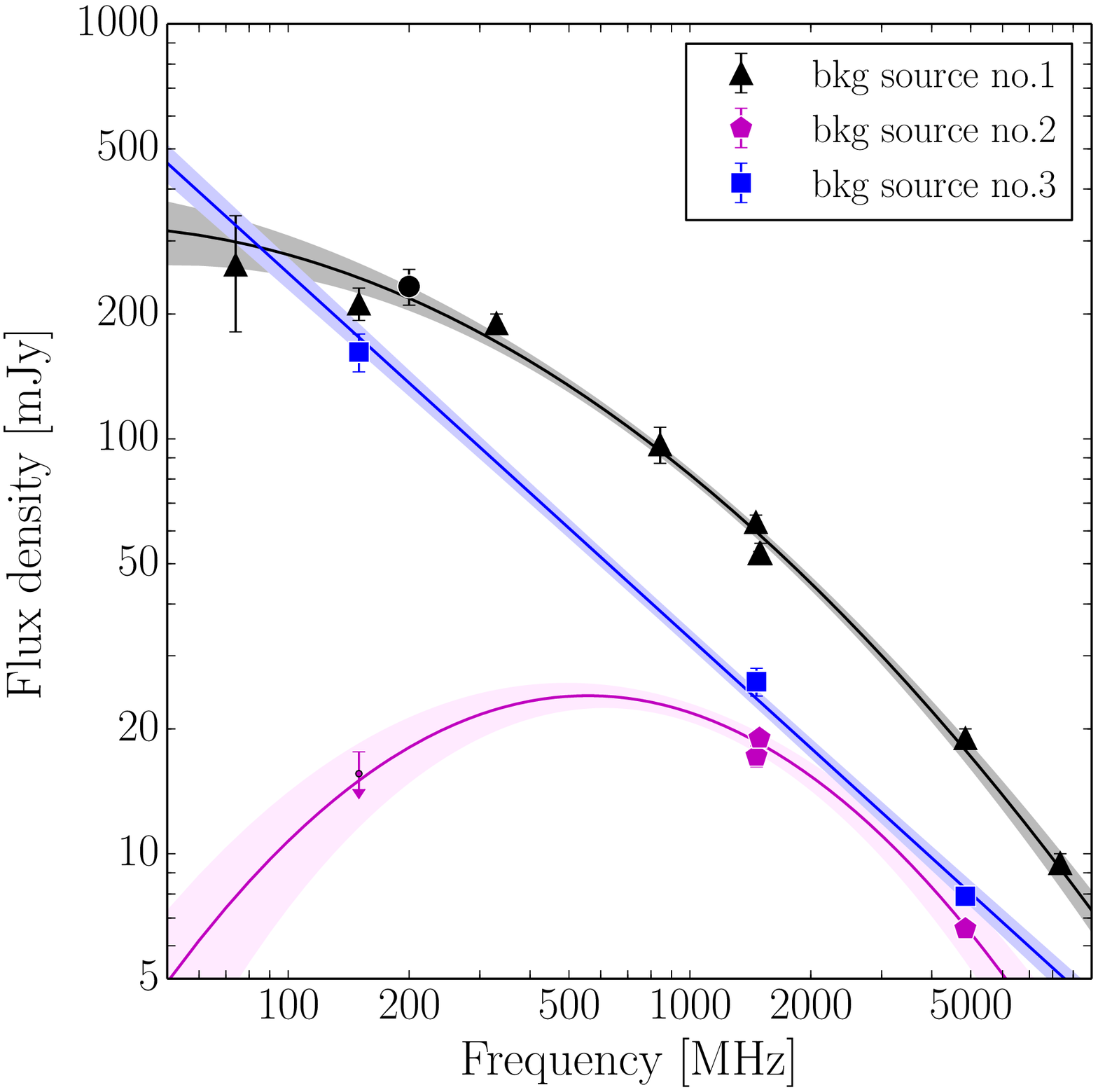}
\caption{{\it Left:} Radio spectra of NGC~253 total radio emission (filled symbols) and central starburst region (empty symbols; measured at an angular resolution of $\sim20-30$~arcsec, equivalent to $\sim300-500$~pc;  {  Table~\ref{tab:spectra-core}}). The measurements from the GLEAM survey are drawn as circles, from the MWA/EoR0 observations as a star and data from literature as triangles (Tables~\ref{tab:spectra}. {  The final spectral energy distribution of the NGC~253 total emission is modeled as a 3-component model (solid line, purple) and is a combination of a power-law and a 2nd order polynomial components modeling the extended emission (drawn cumulatively as dash-dot-dot, green), and internally free-free absorbed synchrotron emission of the central starburst (dash-dot, brown/yellow; refer to \S\ref{sec:spectral-props}). For reference the central starburst modeled as a self-absorbed synchrotron emission (dashed, blue) and 2nd order polynomial (dotted, grey) is also drawn. The flux density contribution of the background sources has been subtracted from the total flux density of NGC~253  (\S\ref{sec:bkg-srcs}).}  {\it Right:} Radio spectra of the background sources. The measurements are listed in Table~{ \ref{tab:spectra-bkg}}, and results on the spectral fitting in Table~\ref{tab:bkgmodels}. The uncertainties on each fit are drawn as shaded areas.}
\label{rys:radio-spectra}
\end{center}
\end{figure*}


\subsubsection{Free-free absorption (FFA)}

The self-absorbed bremstrahlung (i.e. free-free absorbed) radio spectrum can be expressed as \cite[e.g.][]{2002MNRAS.334..912M}
\begin{equation}
S_\nu = S_{\tau=1} \left (\frac{\nu}{\nu_{\tau=1}}\right)^{2} (1 - e^{-\tau_{\rm ff}(\nu)}),
\end{equation}
where the opacity coefficient is given by 
\begin{equation}
\tau_{\rm ff}(\nu) = (\nu/\nu_{\tau=1})^{-2.1}.
\label{eqn:tau-FFA}
\end{equation}

As discussed in \S\ref{sec:intro} radio emission from NGC~253 is a mixture of synchrotron emitting cosmic rays from SNR and thermal emission H{\sc ii} regions. The free-free absorption is expected to start dominating at low radio frequencies, where the intensity of the electrons in the ionized gas becomes high (optically thick regime). For NGC~253 it is a natural assumption that the thermal plasma co-exists with the synchrotron emitting electrons, hence the radio spectrum can be modeled as a synchrotron power-law with an internal free-free absorbing screen \cite[SFA;][]{2003AJ....126..723T},
\begin{equation}
S_\nu = S_{0} \left(\frac{\nu}{\nu_0} \right)^{-\alpha} \left ( \frac{1 - e^{-\tau_{\rm ff}(\nu)}}{\tau_{\rm ff}(\nu)}\right ).
\label{eqn:SFA}
\end{equation}


\subsection{Weighted non-linear least squares fitting}
\label{sec:correlated-noise}

All measurements in this paper are considered independent of each other (in the GLEAM survey valid for flux densities $\gtrsim5$~Jy; see \cite{2016MNRAS.GLEAM.subm}, thus a simple form of $\chi^2$ statistic is used for the fitting of the radio spectra, which at the same time is the goodness-of-fit of the fitted model ($M_i$), 
\begin{equation}
\chi^2 = \sum_{i}^n \left( \frac{\text{M}_i - \text{data}_i}{\text{error}_i} \right)^2
\label{eqn:chi2}
\end{equation}
for $i=1,..,n$ data points. In minimization of Eqn.~\ref{eqn:chi2} we use the Levenberg-Marquardt algorithm \cite[][]{1944Levenberg,1963Marquardt} implemented in the Python\footnote{\tt http://www.python.org} module {\sc lmfit}  \cite[][]{2014PythonLMFIT}.


\subsection{Model selection}
\label{sec:modselct}

For the formal model selection we use the Bayesian inference method. We follow the prescription outlined in \cite{2015ApJ...809..168C}, with the log-likelihood function (the probability of observing the data given model parameters $\theta$) in the form of

\begin{equation}
\text{ln}\mathcal{L}(\theta) = -\frac{1}{2}\sum_{i}^{n}\left[\frac{(\text{M}_i - \text{data}_i)^2}{\text{error}_i^2} + \text{ln}(2\pi\, \text{error}_i^2)\right].
\end{equation}
Under the hypothesis that the models being compared ($M_1$, $M_2$) are equally likely, the model selection can be performed based solely on the Bayesian evidence ($Z$), where

\begin{equation}
\Delta \text{ln}(Z) =  \text{ln}(Z_2) -  \text{ln}(Z_1)
\end{equation}
and
\begin{equation}
Z_{1,2} = \int \int ... \int \mathcal{L}(\theta) \Pi(\theta) d(\theta).
\end{equation}
The dimensionality of the integration depends on the number of model parameters. If  $\Delta\text{ln}(Z)\geqslant3$ model $M_2$ is strongly favoured over $M_1$. If  $1<\Delta\text{ln}(Z)<3$ model $M_2$ is only moderately favoured over $M_1$, and if  $\Delta\text{ln}(Z)<1$ the preference of one model over the other is inconclusive. For more discussion on the theoretical background of the method used see \cite{2015ApJ...809..168C}. We use the {\sc MultiNest} tool \cite[][]{2008MNRAS.384..449F,2009MNRAS.398.1601F,2013arXiv1306.2144F,2014A&A...564A.125B} for our calculations of the Bayesian evidence.


\begin{table*}
  \scriptsize
  \centering
  \caption{Radio spectral indices ($\alpha$) measured for selected regions of NGC~253 (Figure~\ref{rys:sp-idx}). Radio spectral index maps are used at 169~MHz (MWA/EoR0; $S_{\rm 169 MHz}$), 200~MHz (GLEAM; $S_{\rm 200 MHz}$), and 1.465~GHz convolved to the 169~MHz ($S_{\rm 1.4GHz}^{\rm 169 res}$) and separately to 200~MHz ($S_{\rm 1.4GHz}^{\rm 200 res}$) angular resolution. }
  \label{tab:TTplots}
\begin{tabular}{crrrrrrrr}
\hline
\hline
Region & \multicolumn{1}{c}{$\alpha^{169 \rm MHz}_{1.4 \rm GHz}$} & $S_{\rm 169 MHz}$ & $S_{\rm 1.4GHz}^{\rm 169 res}$ & \multicolumn{1}{c}{$\alpha^{200 \rm MHz}_{\rm 1.4 GHz}$} & $S_{\rm 200 MHz}$ & $S_{\rm 1.4GHz}^{\rm 200 res}$ \\ 
       & & \multicolumn{1}{c}{(mJy)} & \multicolumn{1}{c}{(mJy)}  & & \multicolumn{1}{c}{(mJy)} & \multicolumn{1}{c}{(mJy)}\\
\hline
1  & $0.54 \pm0.06$  &  $242\pm26$ &   $75\pm10$ & $0.57\pm0.06$  &  $233\pm25$ &   $75\pm10$\\
2  & $0.63 \pm0.04$  &  $586\pm59$ &  $149\pm13$ & $0.68\pm0.03$  &  $581\pm51$ &  $149\pm13$\\
3  & $0.57 \pm0.02$  &  $883\pm89$ &  $257\pm11$ & $0.62\pm0.02$  &  $884\pm73$ &  $257\pm11$\\
4  & $0.67 \pm0.04$  &  $415\pm42$ &   $97\pm9$  & $0.72\pm0.04$  &  $412\pm36$ &   $97\pm9$\\
5  & $0.31 \pm<0.01$ & $6558\pm656$& $3341\pm14$ & $0.34\pm<0.01$ & $6589\pm528$& $3341\pm14$\\
6  & $0.60 \pm0.07$  &  $224\pm24$ &   $61\pm10$ & $0.70\pm0.07$  &  $246\pm25$ &   $61\pm10$\\
7  & $0.58 \pm0.03$  &  $488\pm49$ &  $139\pm10$ & $0.63\pm0.03$  &  $487\pm42$ &  $139\pm10$\\
8  & $0.50 \pm0.02$  &  $616\pm62$ &   $211\pm9$ & $0.51\pm0.02$  &  $578\pm49$ &  $211\pm9$\\
9  & $0.53 \pm0.06$  &  $216\pm23$ &    $69\pm9$ & $0.51 \pm0.06$ &  $192\pm21$ &   $69\pm9$\\
\hline
\vspace{4mm}
\end{tabular}
\end{table*}


\section{Results}
\label{sec:results}

In what follows we refer to the `halo' as the radio emission beyond the boundary of the optical disk of the galaxy. Given the low angular resolution of the MWA observations we distinguish only between the galaxy disk and the extended synchrotron halo.


\subsection{Background sources}
\label{sec:bkg-srcs}

There are three discrete radio sources located within the extended emission of NGC~253; the sources are marked in Figure~\ref{rys:bkg-sources}, and their positions are based on the NVSS and \cite{2009A&A...494..563H} measurements.

The discrete radio source no.~1 is located at RA(J2000)=$00^h47^m59^s.10$, Dec(J2000)=$-25^\circ18'22''.45$ at 200~MHz, and is most likely a background AGN \cite[][]{1992ApJ...399L..59C}. The radio source is detected in the GLEAM images, but at low frequencies becomes increasingly confused with the NGC~253 halo emission. We measure the flux density of the source only in the deep GLEAM image (Table~\ref{tab:spectra}). The discrete radio source no.~2 is located at RA(J2000)=$00^h47^m12^s.01$, Dec(J2000)=$-25^\circ17'43''.9$ and is most likely a faint background AGN  \cite[][]{2009A&A...494..563H}. This source is heavily embeded in the NGC~253 extended emission in the MWA images. The discrete radio source no.~3 is located at RA(J2000)=$00^h47^m44^s.91$, Dec(J2000)=$-25^\circ13'38''.4$. This source is embeded in the extended emission in our MWA images, but is clearly detected in the TGSS ADR1 image (Figure~\ref{rys:bkg-sources}). The flux density measurements of the background sources are listed in Table~\ref{tab:spectra}, and the spectral modeling results are given in Table~\ref{tab:bkgmodels}. We subtract the estimated flux density contribution of these sources from the total flux density measurements of NGC~253.  In addition, we model the background source no.~1 as a point source with a peak flux density $240$~mJy at 169~MHz and $223$~mJy at 200~MHz, and for pictorial purposes we subtract it directly from the radio image plane. The resulting radio contours are overlaid on H$\alpha$ and X-ray images in Figure~\ref{rys:multi-images} and discussed in \S\ref{sec:halo}.


\subsection{NGC 253}
\label{sec:ngc253}

Radio images of NGC~253 at six chosen radio frequencies are presented in Figure~\ref{rys:radio-images}. At 200~MHz, {  the deepest image from the presented here GLEAM observations,} the size of NGC~253 is 1310~arcsec (major axis) and 535~arcsec (minor axis) measured at a $\text{PA}=52^\circ$, with a total radio luminosity density $2.4(\pm0.1)\times10^{22}$ W Hz$^{-1}$. {  At 169~MHz (MWA/EoR0 image), the size increases by 3--6 percent, to 1440~arcsec (major axis) and 615~arcsec (minor axis), which may be a combination of the intrinsic increase in size and the uncertainty of the measurement}.

\subsubsection{Total radio emission}

Radio continuum spectra of NGC~253 between 76~MHz and 10.7~GHz are plotted in Figure~\ref{rys:radio-spectra} and the flux density measurements are tabulated in Table~\ref{tab:spectra}. Background radio sources (\S\ref{sec:bkg-srcs}) located within the diffuse emission of NGC~253 were subtracted from the total flux density measurements. 
In the construction of the radio spectrum we used archival data provided the measurements were of angular resolution comparable to GLEAM or were sensitive to low brightness emission on angular scales of at least 0.5~deg, the total flux density was integrated over the diffuse emission of NGC~253 and not fitted by Gaussian components, and the absolute flux density scale and the uncertainties of the measurements were quoted. We do not use measurements at angular resolution of $>20$~arcmin because of confusion of NGC~253 with nearby sources. 

We find the best fitting model to be a 2nd-order polynomial with $S_0=7.30\pm0.04$~Jy, $\alpha=0.56\pm0.01$ and a curvature $c_1=-0.12\pm0.01$ at a reference frequency of 1~GHz ($\chi^2=140$, with degrees of freedom: dof~$=45$; Figure~\ref{rys:radio-spectra}), which is significantly preferred to a simple power-law ($\Delta \text{ln}(Z)=100.5\pm0.3$).


\subsubsection{Central starburst region}
\label{sec:nucleus}

The angular resolution of the MWA data is too low to resolve the central starburst region of NGC~253; the highest angular resolution achieved is 102~arcsec at 227~MHz (GLEAM) and 138~arcsec at 169~MHz (EoR), which is over three times the size of the NGC~253 starburst region \cite[][]{1982ApJ...252..102C,1988ApJ...330L..97A,1996A&A...305..402C}. We construct radio spectra of the central starburst region using data from the literature and new measurements from the TGSS ADR1 survey (Figure~\ref{rys:radio-spectra}, Table~\ref{tab:spectra}). We limit the measurements to those that are at an angular resolution comparable to the size of the central starburst region (approximately 20--30~arcsec). 

We find the best fitting model to be a 2nd order polynomial with $S_0=2.28\pm0.02$~Jy, $\alpha=0.20\pm0.01$ and a curvature $c_1=-0.24\pm0.01$ at a reference frequency of 1~GHz ($\chi^2=12.8$, dof~$=13$; Figure~\ref{rys:radio-spectra}). We further attempt to model the spectral turnover, and we find SFA (Eqn.~\ref{eqn:SFA}) to be the best fitting model with $S_{\tau=1}=4.43\pm0.14$~Jy, $\alpha=0.43\pm0.01$ and $\nu_{\tau=1}=238\pm15$~MHz ($\chi^2=42.9$, dof~$=13$). Based on the Bayesian evidence the SFA model (synchrotron plasma absorbed by an internal free-free absorbing screen) is preferred to the pure synchrotron self-absorption model, SSA ($\Delta \text{ln}(Z)=8.3\pm0.3$).


\begin{figure*}
\begin{center}
\includegraphics[width=182mm]{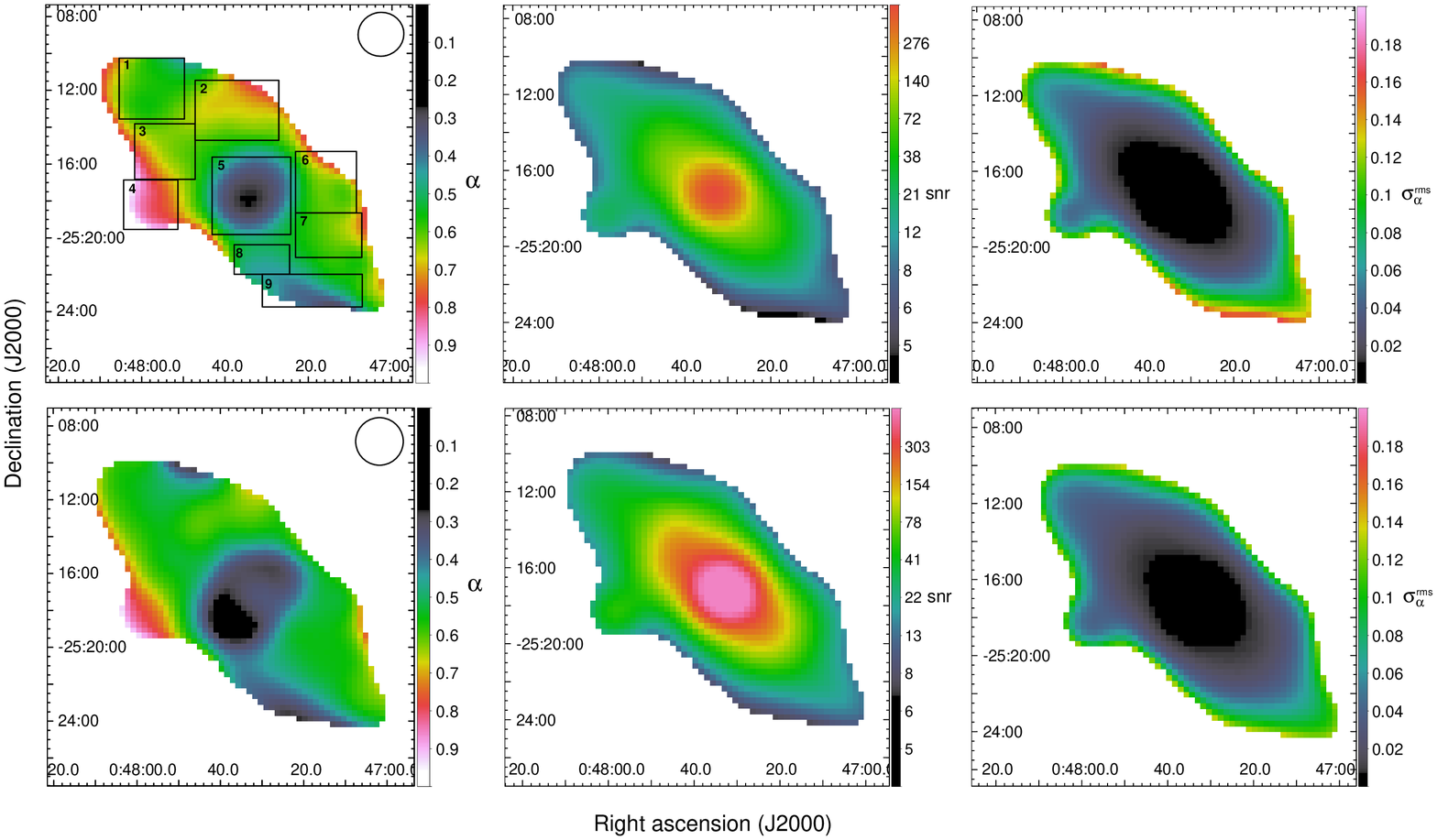}
\caption{{\it Top left:} Radio spectral index distribution map between 200~MHz (GLEAM) and 1.465~GHz \protect\cite[][]{1992ApJ...399L..59C} at an angular resolution of $133\times127$~arcsec$^2$ {  (drawn in the top right corner)}. The color scale is $\alpha$ px$^{-1}$ with the pixel size $18\times18$~arcsec$^2$. The spectral index was calculated only for pixels at $\geq 4\sigma$ rms level at each frequency. We also mark regions for which we calculate their separate spectral index for the T-T method (see \S\ref{sec:spindx-maps}). {\it Top middle:} An average signal-to-noise ratio (snr) map of pixels at 200~MHz and 1.46~GHz used to create spectral index map. {\it Top right:} Uncertainty on the spectral index calculation per pixel based purely on the off source rms noise levels. This uncertainty will be dominant for the low snr pixels. At high {  snr} this uncertainty can only be considered a lower limit. {\it Bottom left, middle, right:} Same as {\it top left, middle, right}, but between 169~MHz (MWA/EoR0) and 1.465~GHz \protect\cite[][]{1992ApJ...399L..59C}, at an angular resolution of $138\times138$~arcsec$^2$.}
\label{rys:sp-idx}
\end{center}
\begin{center}
\includegraphics[width=78mm, angle=270]{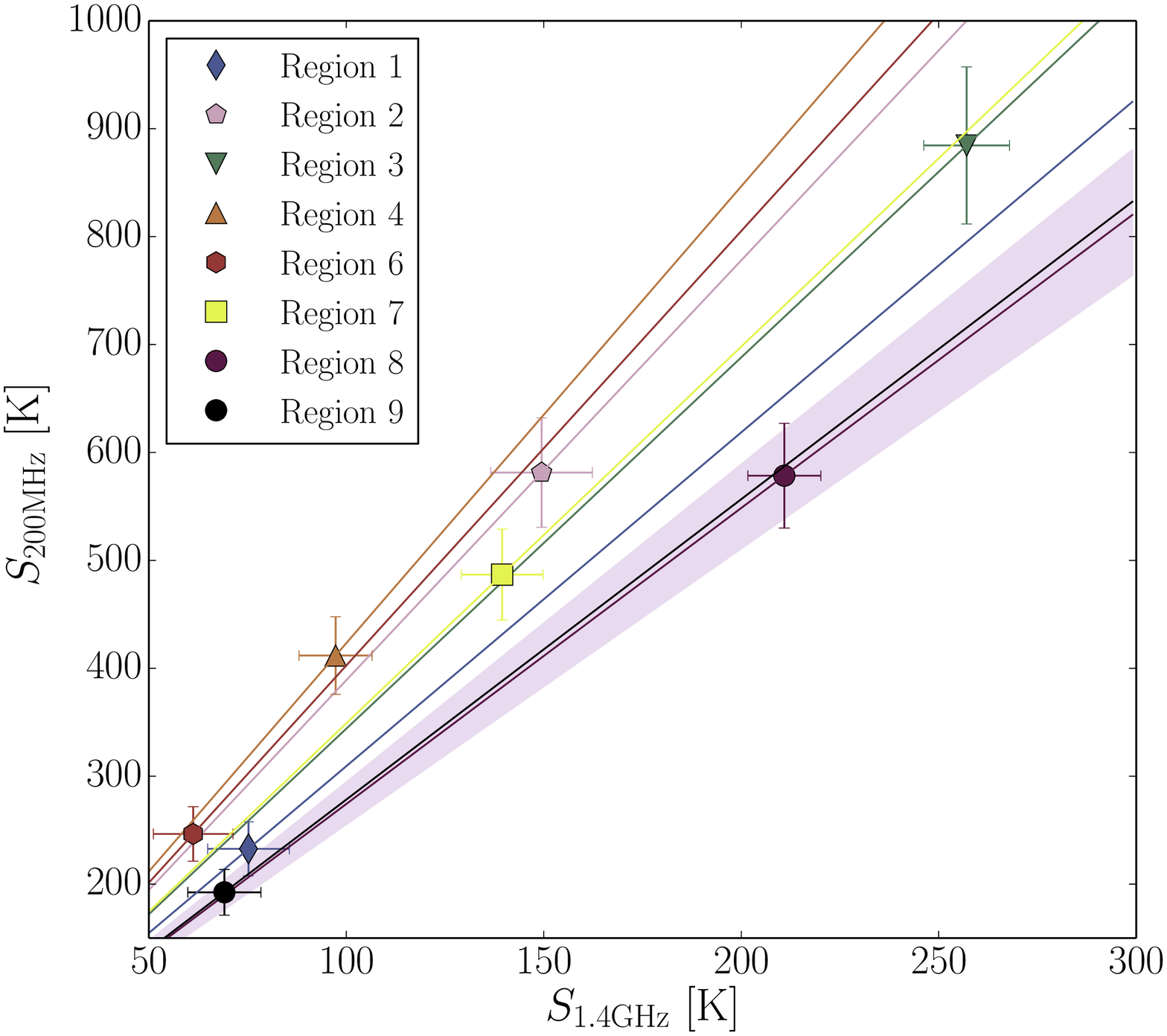}
\includegraphics[width=78mm, angle=270]{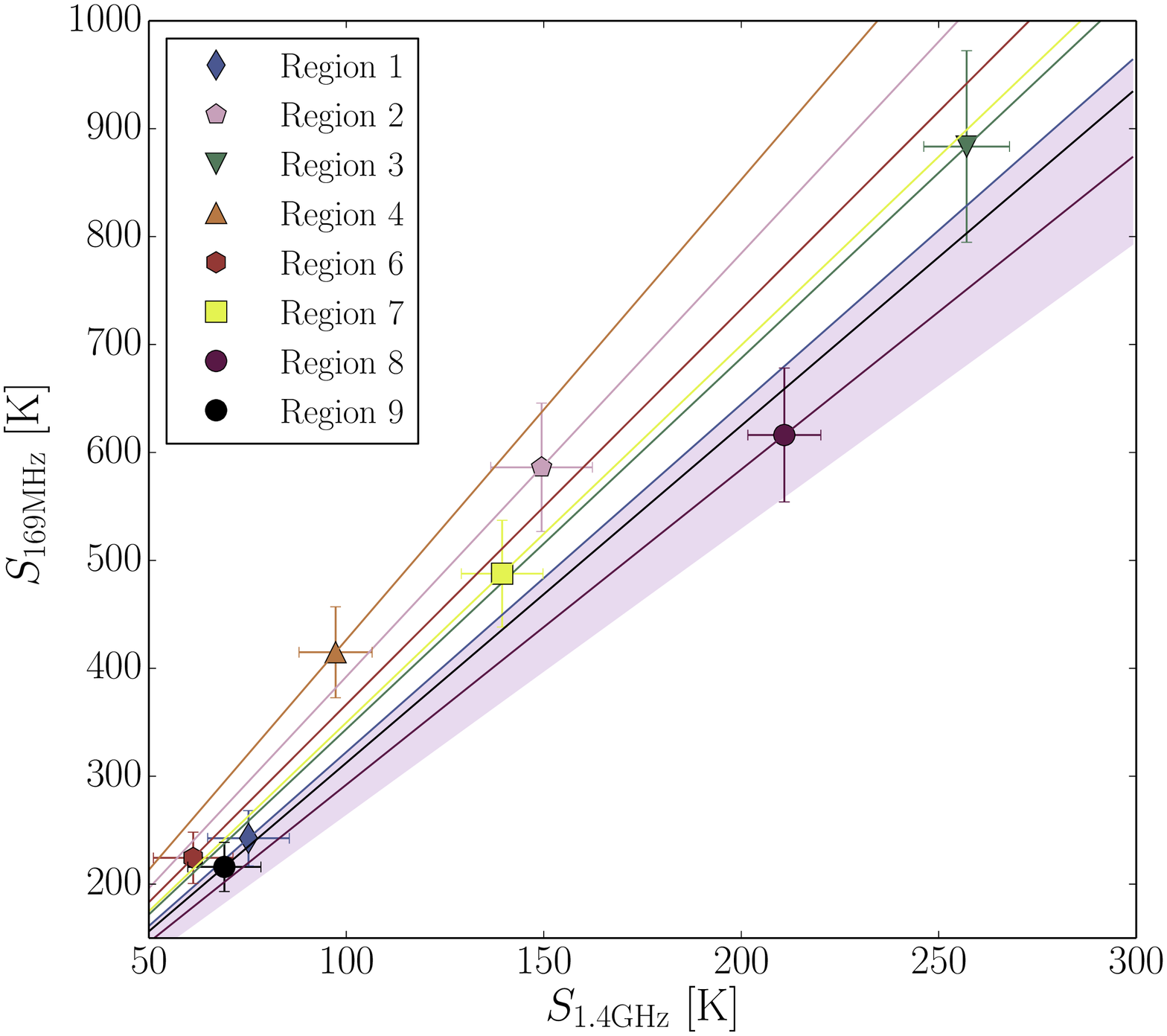}
\caption{T-T plots between 200~MHz and 1.465~GHz ({\it left}) and 169~MHz and 1.465~GHz ({\it right}). Fitted spectral index for each region (Figure~\ref{rys:sp-idx}) is drawn as solid line and is tabulated in Table~\ref{tab:TTplots}. Region 5 (central starburst) is not pictured. For clarity only the uncertainties associated with Region 8 are drawn as a shaded area. }
\label{rys:TTplots}
\end{center}
\end{figure*}


\subsubsection{Spectral index maps}
\label{sec:spindx-maps}

Using the total flux density images at 200~MHz (GLEAM), 169~MHz (MWA/EoR) and 1.46~GHz \cite[][]{1992ApJ...399L..59C}, we created spectral index distribution maps and their corresponding uncertainty and signal-to-noise ratio maps  (Figure~\ref{rys:sp-idx}). {  To create the spectral index maps we convolved the 1.46~GHz image to the resolution of the GLEAM image (for the 200~MHz--1.46~GHz spectral index map), and the MWA/EoR image (169~MHz--1.46~GHz spectral index map).} 
The spectral index maps suggest an apparent variation in $\alpha$ across the NGC~253 disc and halo. The central regions of the galaxy are dominated by a flat component ($\alpha=0.31-0.34$), coinciding with the central starburst. The gradual steepening along the minor axis seen by \cite{2009A&A...494..563H} is seen only on the northern side of the galaxy in our maps. Further, the region extending SW from the central starburst region seems to be flatter ($\alpha \lesssim 0.53$) than in the other parts of the galaxy outside the central starburst ($\alpha \sim 0.60-0.65$).

To verify the significance of the spectral index variation across the galaxy, we used the T-T method \cite[][]{1962MNRAS.124..297T}. The method allows one to estimate a spectral index within defined regions of a source between two frequencies. We define nine regions within NGC~253 (Figure~\ref{rys:sp-idx}). Due to the low angular resolution of our observations and to avoid oversampling, only one data point is associated with each region. The results are shown in Table~\ref{tab:TTplots} and Figure~\ref{rys:TTplots}.

We find that between 200~MHz and 1.465~GHz the apparently flat regions (Region 8 and 9) are statistically different from the other regions within NGC~253 apart from Region 1 (Figure~\ref{rys:TTplots}, Table~\ref{tab:TTplots}). This spectral flattening is not present in the radio spectral index distribution map between 330~MHz and 1.46~GHz of \cite{2009A&A...494..563H}, even though the same high frequency map is used. This clearly indicates that the flattening occurs at $<300$~MHz. There is also a slight flattening of the spectral index in the NE region perpendicular to the major axis (Region 1), although we find this flattening to be statistically different only from Region 2 (eastern NW halo), 4 (radio spur) and 5 (including central starburst) in the $\alpha^{\rm 200 MHz}_{\rm 1.4 GHz}$ map. All regions further flatten at 169~MHz, reducing the differences between spectral indices of the regions. 


\section{Discussion}

\subsection{Low-frequency synchrotron radio halo}
\label{sec:halo}

A large-scale radio halo in NGC~253 was discovered and confirmed by \cite{1992ApJ...399L..59C} and extensively studied by \cite{1994A&A...292..409B}, \cite{2009A&A...494..563H,2009A&A...506.1123H} and \cite{2011A&A...535A..79H}. This synchrotron halo is most pronounced at low radio frequencies, with the estimated scale heights of $1.7\pm0.1$~kpc at 1.4~GHz and $2.5\pm0.2$~kpc at 330~MHz \cite[][]{2009A&A...494..563H}. Both the deep 200~MHz GLEAM image and the MWA/EoR image at 169~MHz reveal the extended synchrotron halo, which is at least as extensive as one detected in the \cite{1992ApJ...399L..59C} 330~MHz map (Figure~\ref{rys:carilli-halo}).


\subsubsection{Maximum vertical extent}

\begin{figure}
\begin{center}
\includegraphics[angle=270,width=88mm]{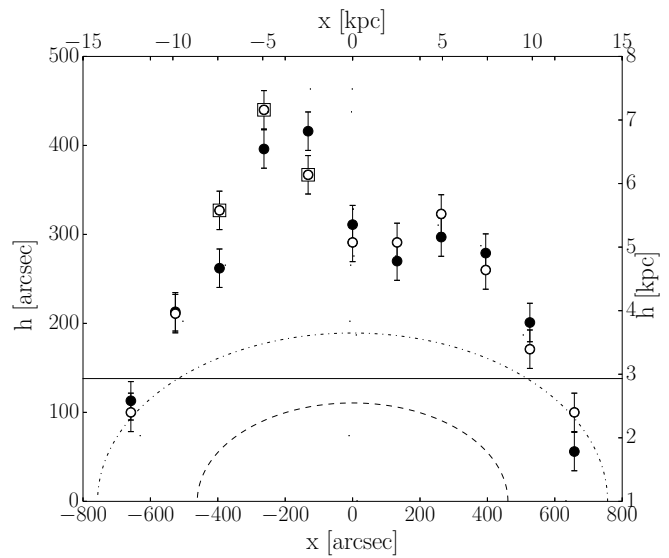}
\caption{ Projected maximum vertical extent of the NGC~253 disk and halo at 169~MHz ($h$) as a function of the distance from the nuclear region along the major axis ($x$), where 0 is centred on the nucleus of the galaxy. The NE direction along the major axis is negative and SW is positive. The extent is are measured perpendicular to the major axis (at $\text{PA} =-38^\circ$) at a step of 132 arcsec (2.4~kpc) separately for the North (filled circles) and South (empty circles) side of the disk and halo as divided by the major axis ($\text{PA} = 52^\circ$). The synthesised beam size of the radio image, $138\times138$~arcsec$^{2}$, is not included in the uncertainties and is drawn as solid horizontal line. Measured sizes of NGC~253 at optical $B$ band including 90~per cent of total light (dash-dotted line) and total infrared $K_{\rm s}$ band (dashed line) are drawn for reference. Plotted values are tabulated in Table~\ref{tab:scaleheights}.}
\label{rys:scaleheights}
\end{center}
\end{figure}

We measured the observed, projected maximum vertical extent of the NGC~253 disk and halo at 169~MHz as a function of the distance from the nuclear region along the major axis (Figure~\ref{rys:scaleheights}, Table~\ref{tab:scaleheights}). The extent is measured perpendicular to the major axis (at $\text{PA} =-38^\circ$, {\it z}-direction) in steps of 132 arcsec (2.4~kpc) separately for the North (filled circles) and South (empty circles) side of the disk and halo as divided by the major axis ($\text{PA} = 52^\circ$). We find the projected radio halo to extend up to 4.75~kpc above the optical \cite[$B$ band including 90~per cent total light,][]{1989spce.book.....L} and 6.3~kpc above the infrared \cite[total $K_{\rm s}$ band,][]{2003AJ....125..525J} edge of the galaxy, reaching up to a total 7.9~kpc in $z$-direction. This is consistent with previous radio measurements at higher radio frequencies \cite[][{  Figure~1}; but cf. Heesen et al. 2009a]{1983PASAu...5..235R,1984A&A...137..138H,1992ApJ...399L..59C}, as well as broadband X-ray observations of the galaxy's extended extraplanar emission \cite[][]{2000A&A...360...24P}. \cite{2009A&A...494..563H,2009A&A...506.1123H} attributed decrease of scale height they measured and modeled to the increased synchrotron losses in the central regions, where the magnetic field is highest.


\subsubsection{Halo morphology}

The shape of the radio halo in our MWA radio images (Figure~\ref{rys:multi-images}) resembles the `horn-like' or `X-shaped' structure seen at GHz radio frequencies \cite[][]{2009A&A...494..563H}, in H{\sc i} \cite[][]{2005A&A...431...65B,2015MNRAS.450.3935L}, X-rays \cite[][]{1988ApJ...330..672F,2000A&A...360...24P,2008AA...489.1029B}, H$\alpha$ (G. Meurer, priv.comm.; Figure~\ref{rys:multi-images}), UV \cite[][]{2005ApJ...619L..99H} and far-IR \cite[][]{2009ApJ...698L.125K}. 

The radio halo was investigated in detail by \cite{2009A&A...506.1123H} who, through modeling of the large-scale magnetic field, attributed its origin to disk wind, confirming previous suggestions \cite[][]{1992ApJ...399L..59C,1994A&A...292..409B}. This is also in line with the H$\alpha$ and optical a\-na\-ly\-ses of the inner starburst-driven superwind \cite[][]{2011MNRAS.414.3719W}. In our MWA maps both the NE and NW halo regions are pronounced. The extended soft X-ray emission ($<1$~keV) of the halo, detected in the north-western direction from the NGC~253 disk, is interpreted as bubbles of hot low density gas \cite[Figure~\ref{rys:multi-images};][]{2000A&A...360...24P,2002ApJ...568..689S}. \cite{2009A&A...506.1123H} postulates that the large-scale magnetic field of the halo follows the walls of these bubbles, where it may be compressed, producing the X-shaped synchrotron radio halo as well as heating up pre-existing cold gas to X-ray energies. The northern halo can also be easily seen, in projection, in our Figure~\ref{rys:multi-images} where we overlay MWA/EoR intensity contours on the soft X-ray emission {\it XMM-Newton} image.


\begin{table}
\caption{ Projected maximum vertical extent of the NGC~253 disk and halo at 169~MHz ($h$) measured at a distance $x$ from the nuclear region along the major axis, where 0 is centred on the nucleus of the galaxy. The NE direction along the major axis is negative and SW is positive. The extent is  measured perpendicular to the major axis (PA$ =-38^\circ$) separately for the North and South side of the disk and halo. The uncertainties on the  measurements are 21.6 arcsec  (equivalent to 0.4~kpc). See \S\ref{sec:halo} for discussion. }
\label{tab:scaleheights}
\begin{center}
\begin{tabular}{rrrrrr}
\hline
\hline 
\multicolumn{1}{c}{$x$}      & \multicolumn{1}{c}{$x$}   & \multicolumn{1}{c}{$h$ North} & \multicolumn{1}{c}{$h$ North} & \multicolumn{1}{c}{$h$ South} & \multicolumn{1}{c}{$h$ South}\\
\multicolumn{1}{c}{(arcsec)} & \multicolumn{1}{c}{(kpc)} & \multicolumn{1}{c}{(arcsec)}  & \multicolumn{1}{c}{ (kpc)   } & \multicolumn{1}{c}{ (arcsec)} & \multicolumn{1}{c}{ (kpc)}\\
\hline
-658  & -11.8  &  113  &  2.0  &  100  &  1.8 \\
-526  &  -9.5  &  213  &  3.8  &  211  &  3.8 \\
-395  &  -7.1  &  262  &  4.7  &  327  &  5.9 \\
-263  &  -4.7  &  396  &  7.1  &  440  &  7.9 \\
-132  &  -2.4  &  416  &  7.5  &  367  &  6.6 \\
0     &   0.0  &  311  &  5.6  &  291  &  5.2 \\
132  &    2.4  &  270  &  4.9  &  291  &  5.2 \\
263  &    4.7  &  297  &  5.5  &  323  &  5.8 \\
395  &    7.1  &  279  &  5.0  &  260  &  4.7 \\
526  &    9.5  &  201  &  3.6  &  171  &  3.1 \\
658  &   11.8  &   56  &  1.0  &  100  &  1.8 \\
\hline
\end{tabular}
\end{center}
\end{table}


The SE region of the extended halo, the `spur' \cite[][]{1992ApJ...399L..59C}, is contaminated by a background source. We modeled the background source as unresolved (at MWA angular resolution) and subtracted it from the 169~MHz EoR and deep 200~MHz GLEAM images as described in \S\ref{sec:bkg-srcs}. The residual emission, which we consider intrinsic to the `spur' is shown in Figure~\ref{rys:multi-images} overplotted on the H$\alpha$ and X-ray images. Although slightly offset \cite[as expected, cf. Fig 19 in][]{2009A&A...506.1123H}, the feature broadly coincides with the extended outflows at both frequencies as clearly seen in our figure. The spur has been previously interpreted as originating from the active star formation region in the NE end of the bar \cite[][]{1988AJ.....95.1057W,1992ApJ...399L..59C}. We confirm the association of the radio `spur' with an extended H$\alpha$ outflow, which can be clearly seen in our Figure~\ref{rys:multi-images} \cite[G. Meurer, priv.comm.;][]{2003PASP..115..928K}. In our radio spectral map between 200~MHz and 1.46~GHz the spur displays the steepest spectral index within the galaxy (Figure~\ref{rys:sp-idx}), gradually steepening from $\alpha\sim0.7$ to $\alpha\sim0.9$ outwards from the galaxy disk (though it still contains the background source no.1 with $\alpha=0.57$), which  is indicative of ageing unabsorbed synchrotron plasma.

\subsection{Spectral properties of NGC~253}
\label{sec:spectral-props}

\subsubsection{Broadband spectrum of total radio emission}
\label{sec:nonthermal-emission}

The broadband spectrum of the total radio emission from NGC~253 is steep, although flattening at MHz radio frequencies (Figure~\ref{rys:radio-spectra}). The total radio emission originates from SNRs, H{\sc ii} regions (predominantly central starburst region; e.g. \citeauthor{1997ApJ...488..621U} \citeyear{1997ApJ...488..621U}, \citeauthor{2000AJ....120..278U} \citeyear{2000AJ....120..278U}, but cf. \citeauthor{1988AJ.....95.1057W} \citeyear{1988AJ.....95.1057W}, \citeauthor{1996AJ....112.1429H} \citeyear{1996AJ....112.1429H}) and electrons (cosmic rays) freely spiralling in the large scale magnetic field \cite[radio halo;][]{2009A&A...494..563H}. The steep spectrum is understood to be of a synchrotron origin. The flattening of the spectrum, however, may be due to a number of reasons, including some degree of absorption of the synchrotron emission (Eqn.~\ref{eqn:SSA}) and a low energy cut-off of the electron population, where in general it is assumed the electron energy ($E$) spectrum can be described as a power-law $N(E)\propto E^{-p}$ with the index $p$ related to the radio spectral index as $\alpha=(p-1)/2$. The low radio frequency flattening of the spectra of starburst galaxies is not unusual and it has been observed previously \cite[e.g.][]{2015AJ....149...32M}.


\begin{figure*}
\begin{center}
\includegraphics[width=170mm]{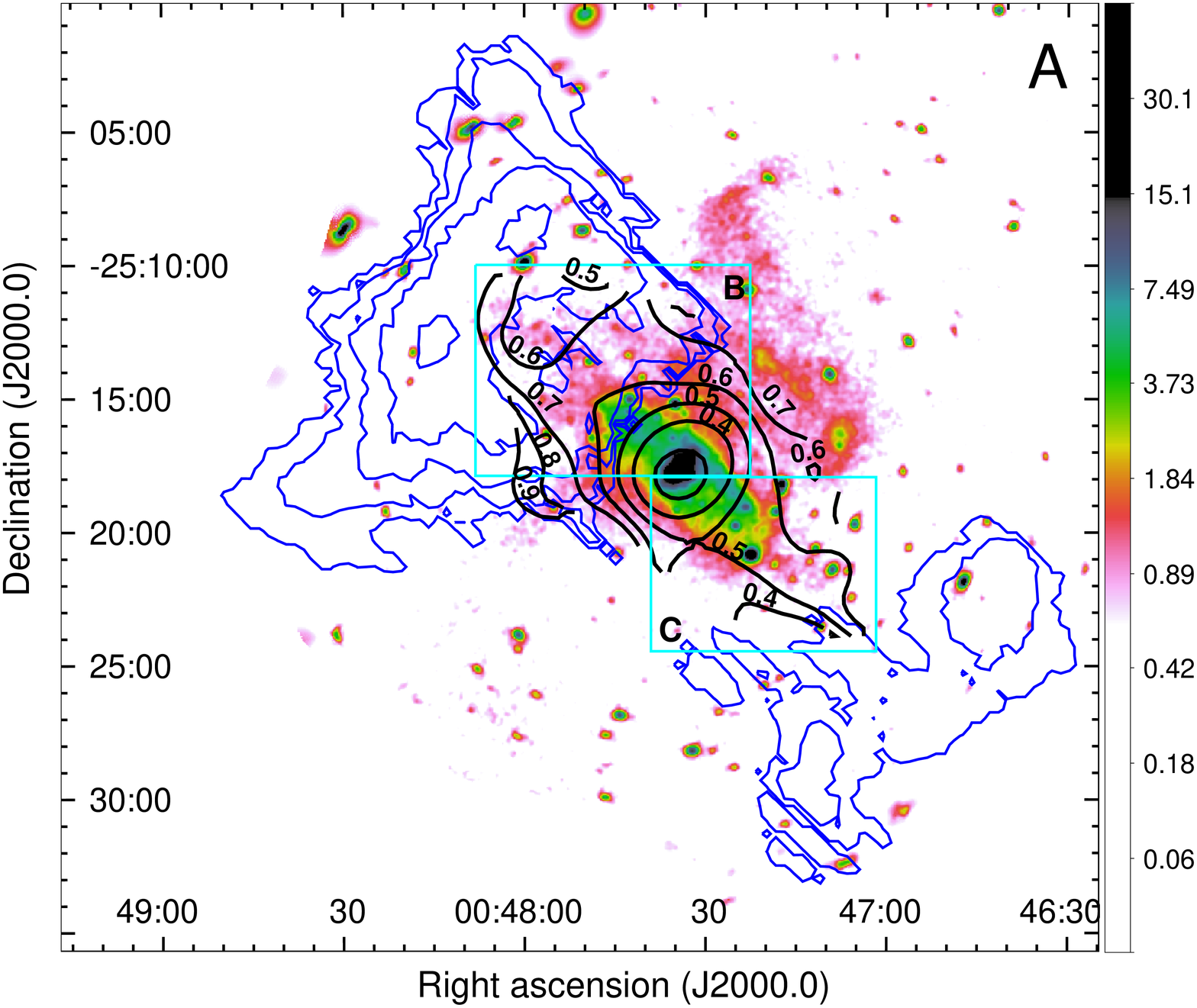}
\includegraphics[width=87mm]{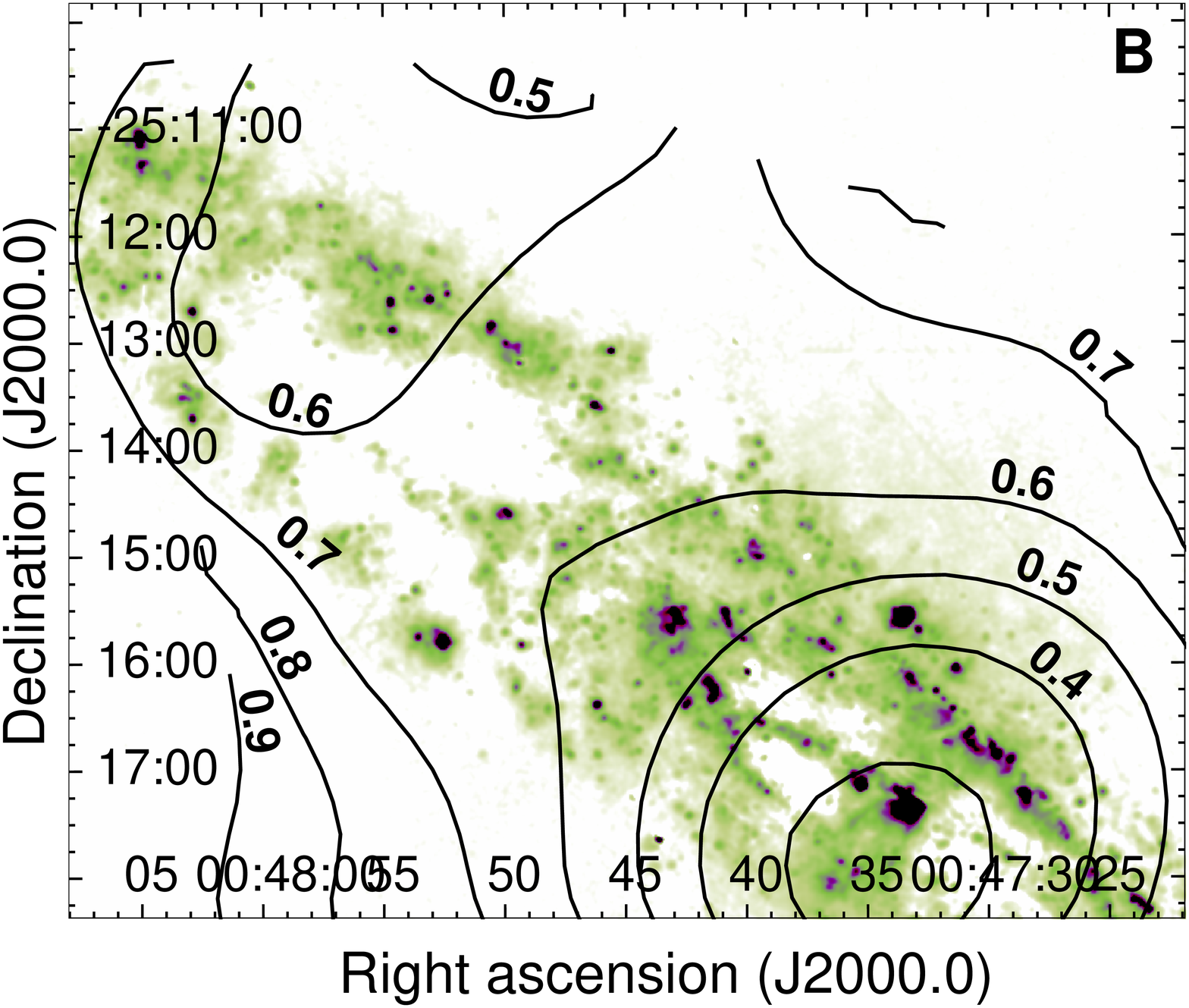}
\includegraphics[width=82.5mm]{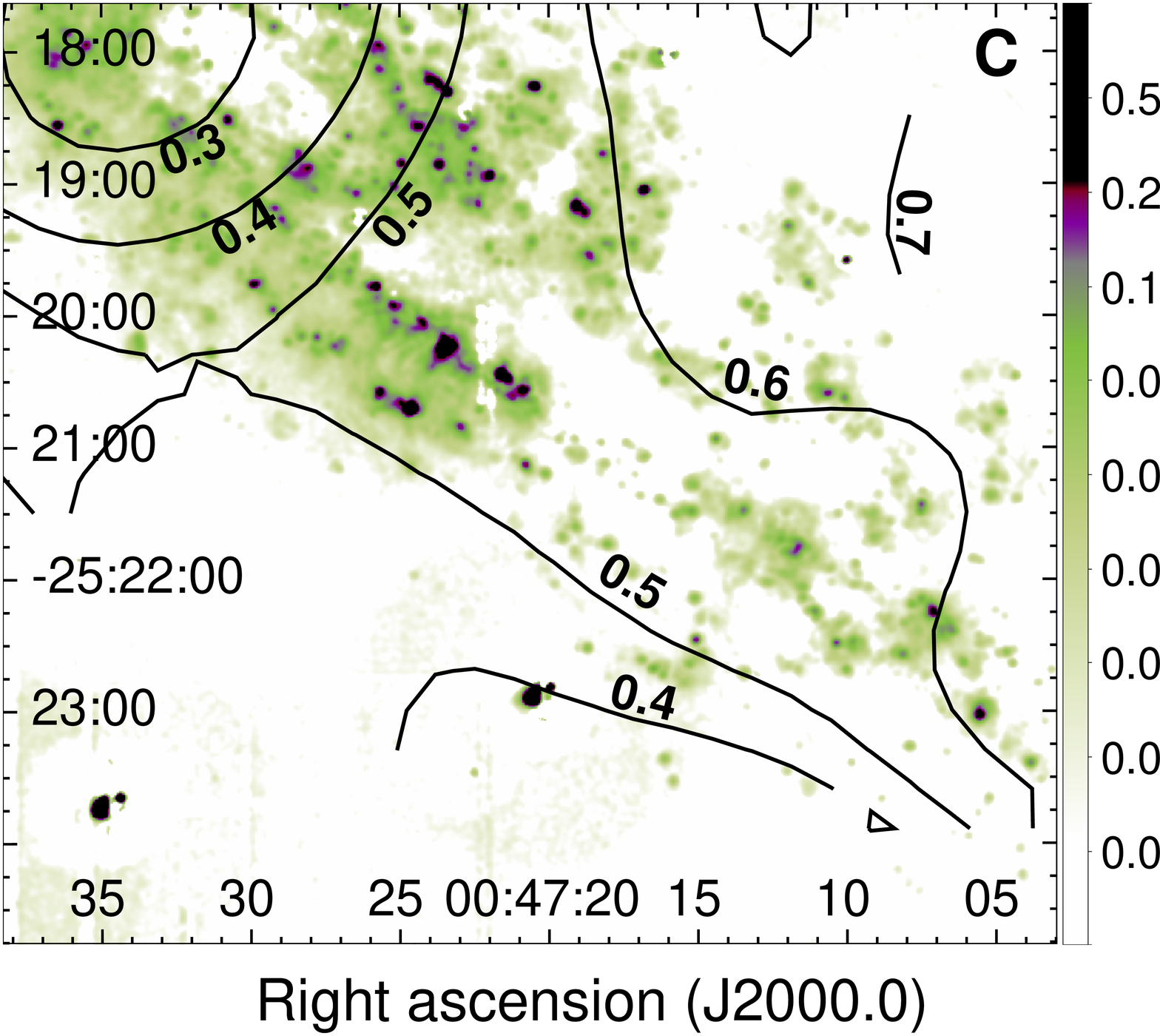}
\vspace{5mm}
\caption{{ (A)} {\it XMM-Newton} image of soft X-ray emission from NGC~253 and its environment \cite[0.2--1 keV band; ][]{2008AA...489.1029B} with overlaid GLEAM radio spectral index distribution (black) and anomalous extraplanar H{\sc i} \protect\cite[][]{2015MNRAS.450.3935L} contours. The color bar is in units of $10^{-5}$~ct~s$^{-1}$~px$^{-1}$. The H{\sc i} contours levels are {  0.09, 0.58, 1.45, 2.9 and 4.8 $\times10^{19}$~cm$^{-2}$.  The beam size of the H{\sc i} observations is $3.5\times3.0$~arcmin$^2$. {  (B) Zoom in on the north disk and NE halo in our H$\alpha$ image (Figure~\ref{rys:multi-images}) with overlaid  GLEAM radio spectral index contours. (C) Zoom in on the SW halo in H$\alpha$ with overlaid  GLEAM radio spectral index contours. The two objects at RA=00:47:20.3, Dec=$-$25:22:51.4 and RA=00:47:34.8, Dec=$-$25:23:41.9 are foreground stars. The colorbar of the panels B and C is in units of ct~s$^{-1}$~px$^{-1}$.}}}
\label{rys:lucero}
\end{center}
\end{figure*}


We attempt to separate the contribution of the extended and central starburst emission to the total flux densities to see if the {  spectral} flattening can be attributed mostly to the absorption occurring in the central starburst region. To  do this we simultaneously fitted the total flux density and the central starburst region spectra, assuming an underlying 2- or 3-component spectrum composed of the central starburst region (component C) and extended emission (component E, composed of one or two components). As an example of our fitting method, in the case where the component E is modeled as a power-law and component C as a free-free absorbed synchrotron emission, the model equation ($S_{\nu}^{\rm mod})$ takes the following form

\begin{equation}
S_{\nu}^{\rm mod} = ( S_\nu^{\rm E}  + S_\nu^{\rm C} )^{\rm tot} + ( S_\nu^{\rm C} )^{\rm cor}, 
\end{equation}
where 
\begin{equation}
S_\nu^{\rm E} = S_0^{\rm ext} \left( \frac{\nu}{\nu_0}\right ) ^{-\alpha^{\rm ext}},
\end{equation}
\begin{equation}
S_\nu^{\rm C} = S_0^{\rm cor} \left( \frac{\nu}{\nu_0}\right ) ^{-\alpha^{\rm cor}} \left (\frac{1-e^{-\tau_{\rm ff}(\nu)}}{\tau_{\rm ff}(\nu)} \right ).
\end{equation}
$\tau_{\rm ff}(\nu)$ is given by Eqn.~\ref{eqn:tau-FFA} and the indices indicate: $\rm tot$ -- total, $\rm cor$ -- core, $\rm C$ -- component C, $\rm E$ -- component E, $\rm ext$ -- extended. This model equation is then compared to our observed data ($S_{\nu}^{\rm obs}$), where $S_{\nu}^{\rm obs} = S_{\nu}^{\rm obs, tot} + S_{\nu}^{\rm obs, cor}$. {  In the 2-component model, we fit the component E with either simple power-law or a curved spectrum (2nd degree polynomial; Eqn.~\ref{eqn:poly}). In the 3-component model, we fit the component E with a combination of power-law and either a curved spectrum, synchrotron self-absorbed (Eqn.~\ref{eqn:SSA})  or synchrotron free-free absorbed component  (Eqn.~\ref{eqn:SFA}). The component C is modeled with either 2nd-degree polynomial, self-absorbed synchrotron or synchrotron power-law emission with a free-free absorbing screen. }

We find the best fitting model to be the 3-component model, with (1) the component E modeled as a combination of a simple power-law and 2nd order polynomial, with a total flux density $S_0 = 5.08\pm0.50$~Jy and $\alpha = 0.71\pm0.01$ at reference frequency 1~GHz, and 34\% of $S_0$ becoming absorbed at low frequencies as described by 2nd order polynomial with $c_1=-0.76\pm0.17$, and (2) the component C modeled as a synchrotron plasma with an internal free-free absorbing screen, with $S_{\tau=1} = 4.34\pm0.11$~Jy, $\nu_{\tau=1} = 231\pm14$~MHz, $\alpha_{\rm SSA} =0.41\pm0.01$. The 3-component model ($\chi^2 = 173$, dof~$=58$) is favoured over any 2-component model, even if the component E is modeled as a 2nd order polynomial ($\Delta\text{ln}(Z)>4.3\pm0.3$). 

Preference of the 3-component model indicates that flattening of the component E spectrum at the lower frequencies is non-negligible. In principle, this flattening could be attributed to synchrotron self-absorption caused by shock re-acceleration of the halo/disk plasma, an external free-free absorbing screen, or an intrinsic low-energy cut-off of the electron distribution. Thermal free-free absorption can be {  largely excluded based on the limited} evidence for high thermal content in the NGC~253 halo, {  especially in the SW region (see \S\ref{sec:spectral-maps} for details)}. Although our Bayesian inference tests indicate that the flattening caused by the synchrotron self-absorption is moderately favoured over the low-energy cut-off in the electron distribution which could be inferred from the 2nd order polynomial fit ($\Delta\text{ln}(Z)=2.5\pm0.3$), we find that the former fit is associated with very high uncertainties. We also find that any model invoking multiple internal components that we tested is strongly favoured over an external free-free absorbing screen. Our results do not change in the absence of the data points between 300 and 600~MHz that may seem unusually high, which further strengthens our result that the radio emitting plasma in the disk and halo of NGC~253 is composed of at least two spectral components that behave differently. This result is also in line with our findings on the radio spectral index variation across the galaxy (discussed further in the next section).

Furthermore, another important result of our spectral modeling is that the central starburst region is best modeled by the SFA model. Although, in principle, the 2nd order polynomial is statistically favoured, the SFA model is more realistic. Curved radio spectra can be explained by low energy cut-off of electron population, SSA, SFA or FFA models. As we have shown the SSA model is statistically ruled out (\S\ref{sec:nucleus}). Given the overwhelming evidence of significant thermal component in the central starburst \cite[e.g. Figure~\ref{rys:multi-images};][]{1997ApJ...488..621U,1999ApJ...518..183K,2011ApJ...739L..24K} coexisting with synchrotron plasma, the synchrotron free-free absorption model is more likely than the low energy cut-off in electron population. The plasma becomes optically thick around frequency $230$~MHz. Given this result, and under a simplified assumption that a uniform optical depth holds across the region, we estimate a typical emission measure \cite[][]{1961RvMP...33..525O,1978ppim.book.....S} of the absorbing gas towards the central starburst region to be very high, of the order $4\times10^5$~pc~cm$^{-6}$ \cite[assuming electron temperature $T_e=4000$~K; cf. discussion in ][]{1996A&A...305..402C}.

\subsubsection{Radio spectral index distribution maps}
\label{sec:spectral-maps}

We now consider the origin of the $\alpha$ variation across the NGC~253 disk and halo. The southern flattening occurs beyond the SW spiral arm, in the halo region. In Figure~\ref{rys:lucero} we overplot the $\alpha^{\rm 200 MHz}_{\rm 1.4 GHz}$ and extraplanar H{\sc i} contours on the {\it XMM-Newton} soft X-ray image of NGC~253. The diffuse X-ray emission indicates ionized hot gas. As pointed out by \cite{2015MNRAS.450.3935L}, the neutral cold H{\sc i} gas seems to surround the X-ray emitting regions. The radio spectral index spatial variations seem to follow the distribution of the X-ray emission, with the $\alpha$ steepening occurring in the regions of intense soft X-ray emission (radio spur and NW halo) and the flattening around the voids of diffuse X-ray plasma (western SE halo, eastern NW halo). This distribution seems to also match the extraplanar H{\sc i} emission, especially in the western SE halo region. {  In H$\alpha$ we detect faint diffuse emission in the NE halo and the southern `spur' (Figure~\ref{rys:multi-images}), with the line fluxes measured down to $3\times 10^{-18}$~erg~s$^{-2}$ s$^{-1}$ arcsec$^{-2}$. In these regions the spectral index seems to steepen (Figure~\ref{rys:lucero}B). There is, however, almost no H$\alpha$ emission present in the western SE halo, while most of the flattening of the component E in the modeling of \S\ref{sec:nonthermal-emission} is due to this region (Figure~\ref{rys:lucero}C, regions 8 and 9 in Figure~\ref{rys:sp-idx}).}

It has previously been suggested that the halo gas originates from both galactic `fountains' from the star-forming disk and a galactic superwind \cite[][]{2000A&A...360...24P,2008AA...489.1029B}. This strong superwind may be pushing and collimating the neutral cold gas in the halo \cite[][]{2015MNRAS.450.3935L}. In the case of strong collimation shocks one may observe flattening of radio spectra due to synchrotron self-absorption in transverse shocks. Our results seem to favour such a scenario. It is also worth noting that the spectrum flattening of the SW halo corresponds to an extended loop, or arch, seen in optical images \cite[blue filter;][]{1982A&A...106..112B}. However, if the SW halo is predominantly diffuse, and of low density, the flattening may be rather due to an intrinsic low-energy cut off of the electron distribution. 

Another important note is that, based on our broadband radio spectrum modeling, the SW halo region cannot be fully responsible for the total spectrum flattening. Radio emission that becomes absorbed at lower frequencies constitutes more than 30\% of the total extended radio emission at 1~GHz ($1.73\pm0.36$~Jy), while the SW region is only 0.77~Jy at that frequency, which means that the flattening must be also occurring, although in a smaller degree, in other regions across the disk and halo.

Although the SW flattening of the radio spectral index is most likely of a synchrotron origin, an external free-free absorbing screen was also previously suggested. The foreground absorption model was favoured by \cite{2008AA...489.1029B} based on their X-ray data modeling and apparent differences of radio and X-ray halo morphologies. As proved later \cite[e.g.][ this publication, \S\ref{sec:halo}]{2009A&A...494..563H, 2009A&A...506.1123H}, deep continuum and polarization radio observations at both GHz and MHz frequencies reveal the horn-like structure of the radio halo, which directly resembles the X-ray diffuse emission. Based on the equipartition assumptions, \cite{2009A&A...494..563H,2011A&A...535A..79H} find the magnetic field within the halo to be very high, $7-12 \mu$G, reaching as much as $160\pm20 \mu$G in the central regions and $46\pm10 \mu$G in the starburst outflow. The magnetic field in the central regions is strong enough for synchrotron emission to contribute a few per~cent to the total X-ray emission \cite[][]{2013ApJ...762...29L}. As discussed in the previous section, we also find that an external free-free absorbing screen is not a statistically preferred model. These new findings support models in which the total X-ray emission may indeed come from a combination of thermal and synchrotron plasma rather than multi-temperature pure thermal plasma with an externally caused absorption \cite[cf.][]{2008AA...489.1029B}.


\section{Conclusions}
\label{sec:conclude}

We present deep, low-frequency radio continuum images and flux density measurements of a nearby, archetypal starburst galaxy, NGC~253. Our data are part of the Galactic and Extragalactic All-Sky MWA Survey and the MWA EoR observations. The images span frequencies between 76 and 231~MHz at angular resolution of 1.7 -- 5 arcmin and rms noise levels of 4 -- 75 mJy (depending on frequency), and present the deepest measurements of NGC~253 at these low radio frequencies yet.

Our main findings are summarized as follows.

\begin{enumerate}

\item We detect a large-scale synchrotron radio halo that at 154--231~MHz displays the X-shaped/horn-like structure seen at GHz radio frequencies, and is broadly consistent with other multiwavelength observations of NGC~253.

\item The projected maximum vertical extent of the synchrotron emission at 169~MHz extends up to 7.5~kpc NW (7.9~kpc SE) from the major axis of NGC~253, consistent with large-scale soft X-ray emission (extending 9~kpc NW) and X-ray outflow (6.3~kpc SE).\\

\item The radio spectrum of the central starburst region of NGC~253 is significantly curved at low radio frequencies, with  the spectral turnover occurring around 230--240~MHz, which is for the first time statistically constrained.

\item The radio spectral index maps show significant spectral variations in the structure of NGC~253 between 200~MHz and 1.465~GHz. In particular, we isolate a region of statistically significant spectral flattening to the western side of the SE halo. However, as the SW region is rather faint at 1.46~GHz it cannot be fully responsible for the total spectrum flattening, which indicates that the flattening must be also occurring, likely in a smaller degree, in other regions across the disk and halo.

\item The broadband spectrum of integrated total radio emission of NGC~253 is best described as a sum of central starburst and extended emission, where the central starburst component is best modeled as an internally free-free absorbed synchrotron plasma, and the extended emission as synchrotron emission flattening at low radio frequencies. We also find that an external free-free absorbing screen is not a statistically preferred model when compared to models including multiple internal components.

\item We find that the extended emission of NGC~253 is best modeled by a combination of two synchrotron components, one of which becomes significantly absorbed at low radio frequencies. The flattening occurs at frequencies below $300$~MHz, and may be attributed to synchrotron self-absorption of shock re-accelerated electrons or an intrinsic low-energy cut off of the electron distribution.

\end{enumerate}


\section*{Acknowledgments}

ADK thanks P.~A. Curran for valuable discussions on data modeling and constant encouragement in achieving the goals. The authors thank the anonymous referee for careful reading of the manuscript and suggestions that improved this paper. The authors thank W.~Pietsch and D.~Lucero for providing, respectively, X-ray and H{\sc i} fits images of NGC~253, and O.I. Wong and X. Sun for helpful comments. The authors thank V. Heesen for 1.465~GHz image of NGC~253 and for helpful discussions. SB acknowledges funding for the ICRAR Summer Scholarship. 

This research was conducted under financial support of the Australian Research Council Centre of Excellence for All-sky Astrophysics (CAASTRO), through project number CE110001020. 
This scientific work makes use of the Murchison Radio-astronomy Observatory, operated by CSIRO. We acknowledge the Wajarri Yamatji people as the traditional owners of the Observatory site. Support for the operation of the MWA is provided by the Australian Government (NCRIS), under a contract to Curtin University administered by Astronomy Australia Limited. We acknowledge the Pawsey Supercomputing Centre which is supported by the Western Australian and Australian Governments.
This research has made use of the NASA/IPAC Extragalactic Database (NED) which is operated by the Jet Propulsion Laboratory, California Institute of Technology, under contract with the National Aeronautics and Space Administration.
This publication uses the following radio data reduction software: the Multichanel Image Reconstruction, Image Analysis and Display software \cite[{\sc{Miriad}};][]{1995ASPC...77..433S}, the Common Astronomy Software Applications package \cite[{\sc {CASA}};][]{2007ASPC..376..127M} and the Astronomical Image Processing System {\sc AIPS}. {\sc AIPS} is produced and maintained by the National Radio Astronomy Observatory, a facility of the National Science Foundation operated under cooperative agreement by Associated Universities, Inc.

\label{lastpage}

\end{document}